\documentclass[aps,prb,twocolumn,superscriptaddress,nofootinbib]{revtex4-2}

\usepackage{amsmath,amssymb,amsfonts,bm}
\usepackage{graphicx}
\usepackage{xcolor}
\usepackage[colorlinks=true,
            linkcolor=blue,
            citecolor=blue,
            urlcolor=blue]{hyperref}

\renewcommand{\vec}[1]{{\boldsymbol #1}}

\begin{document}

\title{Tomographic imaging of superconducting order using particle-hole interference}

\author{Archisman Panigrahi}
\email{archi137@mit.edu}
\affiliation{Department of Physics, Massachusetts Institute of Technology, Cambridge, MA 02139}

\author{Vladislav Poliakov}
\affiliation{Department of Physics, Massachusetts Institute of Technology, Cambridge, MA 02139}

\author{Leonid Levitov}
\affiliation{Department of Physics, Massachusetts Institute of Technology, Cambridge, MA 02139}

\date{\today}

\begin{abstract}
Superconducting phases with exotic symmetries that differ from the underlying crystalline lattice are at the focus of superconductivity research. Yet, despite intense interest, detecting the order parameter symmetry and topology remains a major challenge. Real-space imaging near atomic impurities with scanning tunneling microscopy (STM) has been highly successful in revealing nodes of the superconducting gap, in particular in cuprate superconductors, however the order parameter phase winding has so far remained inaccessible by STM techniques. We demonstrate that STM can access this phase information by exploiting Young-type quasiparticle interference patterns generated by pairs of impurities acting as beam splitters. Superconducting order parameter tomography (SOPT), a technique proposed here, utilizes the response of real-space interference patterns of Bogoliubov quasiparticles to the controlled rotation of impurity configurations, allowing us to reconstruct the momentum space structure of the gap function $\Delta(\vec{k})$. As a concrete example, we consider Strontium Ruthenate, whose superconducting order remains a subject of ongoing debate, and demonstrate how SOPT can distinguish between competing order parameter candidates. The Young's interference fringes, nodal directions, and rotating beams, detected by SOPT, encode information about both the nodes and phase winding of the superconducting order parameter. This method provides a broadly applicable route to identifying unconventional and topological superconductivity and establishes particle-hole interference as a new imaging modality for superconducting order.
\end{abstract}

\maketitle

\section{Introduction}

Mapping superconducting pairing on the Fermi surface is a central goal in superconductivity research. Of particular interest are unconventional and topological superconductors, where the gap function $\Delta(\vec{k})$ exhibits phase winding when time-reversal symmetry is broken, and nodal structures when it is preserved. Experimental techniques for probing $\Delta(\vec{k})$ fall broadly into two categories: those operating in momentum space---such as ARPES \cite{ref1,ref2,ref3,ref4}, thermal transport \cite{ref5,ref6}, ultrasound \cite{ref7}, and neutron scattering \cite{ref8}---and those operating in real space, including tunneling \cite{ref9,ref10,ref11,ref12}, and near-field optical or magnetic probes \cite{ref13,ref14,ref15,ref16,ref17}. A similar dichotomy appears in theoretical work. In $\vec{k}$-space, approaches to detect exotic superconductivity focus on Raman scattering, anisotropic heat transport, and neutron scattering \cite{ref18,ref19,ref20,ref21,ref22,ref23,ref24,ref25,ref26,ref27,ref28}, whereas real-space approaches focus on tunneling spectroscopy and Yu-Shiba-Rusinov and majorana bound states \cite{ref29,ref30,ref31,ref32,ref33,ref34,ref35,ref36,ref37,ref38,ref39,ref40,ref41,ref42,ref43,ref44,ref45}.

\begin{figure}[h]
\centering
\includegraphics[width=1.0\linewidth]{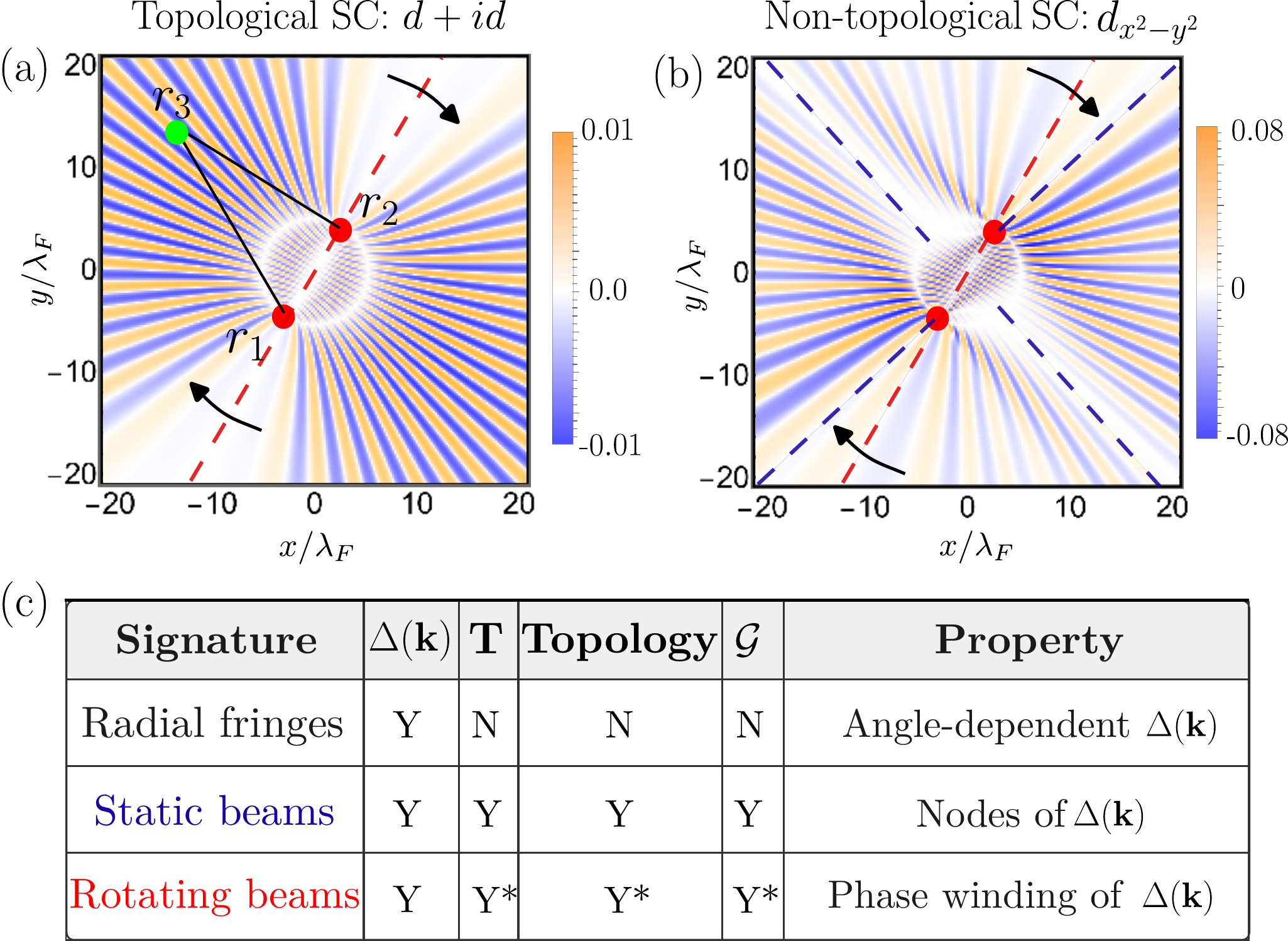}
\caption{
(a and b) Signatures of unconventional superconductivity: particle-hole contribution to the tunneling density of states at zero bias, TDOS at $eV\ll k_B T$, Eq.~\eqref{eq:tdos}, near two impurities at positions $\vec{r}_1,\vec{r}_2$, displaying qp interference patterns as a function of the STM tip position $\vec{r}_3$. The interference pattern changes as a response to the rotation (black arrows) of the impurities, featuring static and rotating nodal beams, which capture information about the symmetry properties of the superconducting gap function $\Delta(\vec{k})$. The red lines represent rotating nodal beams that rotate along with the impurities, reflecting the angular dependence and phase winding of $\Delta(\vec{k})$, Eq.~\eqref{eq:visibility}. In contrast, blue lines mark static beams associated with nodes of $\Delta(\vec{k})$, which remain pinned to crystal axes. The calculations assume a circular Fermi surface with $\xi=10^3\lambda_F$, impurity separation $|\vec{r}_1-\vec{r}_2|=10\lambda_F$, and temperature $k_B T=0.1\Delta$. The plotted quantity is measured in the units of $4e^2|t_{\rm STM}|^2\nu_0 U_0^2\left(k_F/2\pi\lambda_F v_F^2\right)^{3/2}$. (c) A summary of telltale signatures for $\Delta(\vec{k})$ for different orders. Here Y and N indicate whether different observable signatures in the qp interference imply symmetry properties of the angle-dependent gap function $\Delta(\vec{k})$ (for details see text beneath Eq.~\eqref{eq:bogoliubov_wavefunction}).
}
\label{fig:fig1}
\end{figure}

Although these theoretical and experimental methods have led to major advances, progress is still constrained by a key trade-off: Probes sensitive to $\vec{k}$-space information tend to average over $\vec{r}$-space, and vice versa. Many authors have noted that simultaneous access to both spaces would provide crucial insight. In particular, STM studies have shown that real-space measurements can reveal features of the momentum-space structure of $\Delta(\vec{k})$, including the sign changes of a real gap function \cite{ref29,ref30,ref31,ref32,ref33, ref38,ref39,ref42,ref43,ref44}. However, a general framework connecting $\vec{r}$- and $\vec{k}$-space observables to the symmetry and complex phase winding of $\Delta(\vec{k})$ has been lacking, as the complex phase cancels out in traditional experiments such as the quasiparticle interference (QPI) experiment around an impurity. It is natural to ask whether the complex phase winding in a topological superconductor can be detected with a real-space technique.

To answer this question and address this gap, here we introduce a ``superconducting order parameter tomography'' (SOPT) approach. Tomography is a method that reconstructs an object's internal structure from measurements taken from different angles. In the SOPT method, we rotate the pair of impurities and use STM maps to extract the complex phase winding and the node(s) of $\Delta(\vec{k})$ on the Fermi surface. We analyze tunneling density of states (TDOS) $dI/dV$ patterns near two impurities (Fig.~\ref{fig:fig1}) that were recently predicted to exhibit hyperbolic interference fringes \cite{ref44}. We identify two types of features in these patterns that distinguish topological from nontopological superconductivity. Namely, as the impurity pair is rotated, interference vanishes along specific directions, forming static and rotating nodal beams: static beams arise because the Young's interference pattern vanishes along the gap $\Delta(\vec{k})$ nodes, while the rotating nodal beams encode the phase winding, i.e. the superconducting state topology. Importantly, these features are generic and occur for realistic band structures (see below, Figs.~\ref{fig:fig4} and \ref{fig:fig6}).

The static and rotating beams in QPI patterns in TDOS maps offer a diagnostic of key properties such as $\Delta(\vec{k})$ angular dependence, phase winding (topology), time-reversal symmetry $(T)$, and discrete symmetries of the point group $\mathcal G$. The QPI effects arise from Bogoliubov quasiparticles---coherent superpositions of counterpropagating particle and hole states (Fig.~\ref{fig:fig2}) with phases of opposite signs as they propagate and scatter \cite{ref44,ref46,ref47,ref48,ref49,ref50,ref51,ref52,ref53},
\begin{equation}
\Psi_\alpha(\vec{r})
=
\sum_{\vec{k}} a_{\vec{k}}
\left(
u_{\vec{k}} e^{i\vec{k}\cdot\vec{r}} c_{\vec{k},\alpha}
+
\eta_{\alpha\beta}
v_{-\vec{k}} e^{-i\vec{k}\cdot\vec{r}}
c^\dagger_{-\vec{k},\beta}
\right),
\label{eq:bogoliubov_wavefunction}
\end{equation}
where $\alpha$ and $\beta$ are spin indices. The tensor $\eta_{\alpha\beta}$ describes the paired state (with $\eta_{\alpha\beta}=i\sigma^y_{\alpha\beta}$ for spin-singlet pairing). The coefficients $a_{\vec{k}}$ depend on the impurity potential. The spatial interference patterns in the TDOS measured near two impurities arise due to the opposite phase factors of electrons and holes in Eq.~\eqref{eq:bogoliubov_wavefunction}. The key features of these patterns and their relation to system symmetries are summarized in the table in Fig. \ref{fig:fig1}(c):

Radial fringes arise due to particle-hole interference when the gap function $\Delta(\vec{k})$ modulus and phase are strongly angle dependent and are absent when $\Delta$ is constant. Yet, without additional features such as static or rotating beams, radial fringes alone do not reveal information about time-reversal symmetry, topology, or discrete symmetries.  This is reflected by the row Y N N N.

Static nodal beams align with nodal directions of the gap function $\Delta(\vec{k})$, along which coherent particle-hole interference vanishes. If a beam aligns with a lattice symmetry axis, it signals odd parity under the corresponding mirror symmetry.  In general, the presence of static beams due to gap function nodes signals unbroken TRS. This is reflected in the row YYYY.

Rotating beams reflect phase winding; its value can be read from the number of rotating nodal beams. Namely, phase winding $4\pi n$ translates into $2n$ distinct rotating nodal beams, see Eq.~\eqref{eq:visibility} and discussion beneath it. This relation can be more complicated when the modulus $|\Delta(\vec{k})|$ is not nearly uniform but has strong angular dependence (e.g., see Fig. \ref{fig:fig1}(b)), reflected by asterisks in the row Y Y* Y* Y*.

An experimental approach relying on two-impurity interference would proceed as follows: First, one needs to identify or prepare two impurities in close proximity, separated by several Fermi wavelengths or less, with other impurities being sufficiently far away. It is possible to find such a configuration when the impurity concentration is small. Alternatively, two impurities can be placed on an otherwise clean sample using STM. The tunneling density of states $dI/dV$ measured in the surrounding region encodes the information about the QPI effect, however it can be ``overshadowed'' by Friedel oscillations’ contributions. To suppress the Friedel oscillations' effects and make the QPI stand out, the measured TDOS pattern can be Fourier filtered. For a particular material, the wavenumbers corresponding to Friedel oscillations are usually well known,---from, e.g., the Fourier Transform Scanning Tunneling Spectroscopic (FT-STS) measurement \cite{ref55}. After performing inverse Fourier transform the resulting spatial map will be dominated by Young's interference patterns contributed by two impurities. These patterns, in particular the features that follow the impurities when they are rotated (or remain stationary), can be used to identify the signatures of superconducting order illustrated in Fig.~\ref{fig:fig1}, summarized in the table and discussed below. Since the intensity of Young's interference pattern only depends on the product of the strengths of the two impurities, the method does not require the two impurities to be identical or have identical strengths.

\section{Qualitative Picture}

In this section we will discuss the spatial structure of two-impurity interference. The spatial structure hosts not only Young’s interference fringes, but also static and rotating nodal beams introduced above. Now we discuss why Young’s interference fringes appear in the TDOS near two impurities, and why the QPI pattern hosts such nodal beams. STM studies near isolated impurities is a popular measurement that can greatly enrich our understanding of the properties of the pair state \cite{ref9,ref10,ref11,ref12,ref34,ref35,ref43}. Multiple impurities can act as beam splitters, and can further enrich this picture by inducing quasiparticle interference, which manifests as fans of fringes in the TDOS \cite{ref44}.

This behavior is simplest to understand for a toy model of a two-dimensional Fermi gas with a parabolic band and weak-coupling superconductivity, described by an angle-dependent order parameter $\Delta(\vec{k})$. For two impurities, the quasiparticle interference effect---reminiscent of Young's double-slit experiment---contributes to the TDOS as,
\begin{equation}
\frac{dI}{dV}(V,\vec{r}_3)
\sim
\cos k_F(r_{31}-r_{32})\times \ldots,
\label{eq:young}
\end{equation}
where $r_{31}$ and $r_{32}$ are the distances from the STM tip ($\vec{r}_3$) to the two impurities ($\vec{r}_1$ and $\vec{r}_2$), and the dots indicate distance-dependent terms discussed below. This effect is distinct from Friedel oscillations induced by one impurity, which, at distances $r\lesssim\xi$ and nonzero temperature, are similar in the normal and superconducting states. The quasiparticle interference arises in Eq.~\eqref{eq:young} because a quasiparticle can travel as an electron-like state, $e^{ik_F r_{31}}$, from the tip to one impurity, and as a hole-like state, $e^{-ik_F r_{23}}$, from the other impurity to the tip. The resulting interference fringes are hyperbolas, with the total number approximately equal to the integer nearest to $2r_{12}/\lambda_F$. Such fringes are a generic feature of the superconducting state---largely insensitive to pairing symmetry---and are absent in the normal state as there the electrons and holes do not interconvert.

However, the interference patterns also exhibit additional features---namely, static and rotating nodal beams. The beams represent radial regions in which the fringes exhibit $\pi$ jumps, in which their visibility is diminished. We illustrate the properties of beams using chiral $d+id$ and nonchiral $d_{x^2-y^2}$ gap functions,
\begin{align}
{\rm a)}\quad
\Delta(\vec{k})=
|\Delta|e^{i\theta_{\vec{k}}}
\sim
(k_x+ik_y)^2;
{\rm b)}\quad
\Delta(\vec{k})\sim
k_x^2-k_y^2,
\label{eq:gaps}
\end{align}
$|\vec{k}|\approx k_F$, which produce qp interference patterns shown in Fig.~\ref{fig:fig1}. As we will see, the beams are sensitive to the angular structure and symmetry of the paired state.

For the $d+id$ state, the interference pattern features rotating nodal beams aligned with the impurities. These beams rotate when the impurities are moved (see red lines in Fig. \ref{fig:fig1}(a)). As discussed below, this behavior is tied to the nontrivial phase structure of $\Delta(\vec{k})$. In contrast, the $d_{x^2-y^2}$ state displays both rotating and static nodal beams. The rotating beams behave similarly to those in the $d+id$ case. The static beams, however, remain fixed at $\pm45^\circ$ and $\pm135^\circ$, regardless of impurity orientation. These are associated with nodal lines of $\Delta(\vec{k})$, well known from the studies of cuprates \cite{ref56}. Because nodal directions of $\Delta(\vec{k})$ are fixed in momentum space, the corresponding interference beams retain fixed orientation in real space (Fig. \ref{fig:fig1}(b)). When static beams are absent and the superconducting gap has nearly uniform magnitude, rotating beams capture information about phase winding of $\Delta(\vec{k})$. Rotating beams and their connection to the topological phase of the gap function, to the best of our knowledge, have not been previously discussed.

Static and rotating nodal beams in the interference pattern reflect $\Delta(\vec{k})$ symmetry and phase winding. Indeed, as we show below, the TDOS in a spin-singlet superconductor is proportional to
\begin{equation}
\frac{dI}{dV}(V,\vec{r}_3)
\sim
{\rm Re}\left[
\Delta^*(\vec{k}_3)
\left(
\Delta(\vec{k}_3)-\Delta(\vec{k}_{12})
\right)
\right]\ldots,
\label{eq:visibility}
\end{equation}
where $\vec{k}_{12}$ and $\vec{k}_3$ are wavevectors pointing along the line $\vec{r}_1-\vec{r}_2$ and from the midpoint $(\vec r_1+\vec r_2)/2$ to the tip. Eq.~\eqref{eq:visibility} holds for a singlet superconductor, in which $\Delta(\vec{k})=\Delta(-\vec{k})$; a similar result holds for a triplet superconductor. This makes TDOS sensitive to the angular dependence of the gap function, allowing phase information to be inferred from the structure of the interference fringes. In the semiclassical limit, a quasiparticle traveling between two points $\vec{r}_i$ and $\vec{r}_j$ has momentum $\vec{k}_{ij}=k_F\hat{\vec{r}}_{ij}$. There are special directions where this contribution vanishes, defining rotating nodal beams. In a topological superconductor, Eq.~\eqref{eq:gaps}a, the resulting pattern only depends on relative phase difference and thus rotates uniformly. This behavior persists even when $|\Delta(\vec{k})|$ varies moderately. In this case, a spin-singlet topological superconductor with uniform magnitude of the gap and phase winding $4n\pi$ will have $2n$ distinct rotating nodal beams.  Importantly, the relation between the nodal beams' number and the gap function phase winding does not depend on the distance between impurities or on their orientation. 

\section{Practical Considerations}

Here we consider ways to single out the two-impurity interference effect, making it visible despite stronger Friedel oscillations from individual impurities. The interference fringes have distinct characteristics that allow for isolation and analysis. In some cases, such as the Sr$_2$RuO$_4$ system discussed below, the square Fermi surface shape makes the two-impurity interference stand out. In other cases, such as the circular-symmetric toy model discussed above, the QPI fringes occur in the same region in space as Friedel oscillations and may be challenging to see directly. However, even in this case, the low-visibility QPI fringes can be revealed by Fourier filtering. By filtering out components with wavenumbers $k\approx 2k_F$ and transforming back to real space, one can isolate the two-impurity interference contribution that has characteristic wavenumbers parametrically smaller than $2k_F$. These two approaches, which allow to directly access the QPI effect, are illustrated in the analysis of Sr$_2$RuO$_4$ below and in the discussion of Fourier filtering (see ref. \cite{ref81} and SI Appendix, section F).

Another consideration has to do with optimizing the system. The two-impurity interferometry experiment sensitive to $\Delta(\vec{k})$ phase will benefit from a small number of Fermi-surface sheets and strongly angle-dependent $\Delta(\vec{k})$. In that regard, superconductors in the FeSe family (hosting highly anisotropic yet nodeless gaps) and Bi$_2$Sr$_2$CaCu$_2$O$_{8+\delta}$ (strongly anisotropic and sign-changing gaps) are particularly appealing, as well as the unconventional, and possibly topological superconductors Sr$_2$RuO$_4$ and UTe$_2$.

\section{Microscopic Mechanism}

To establish these results, we consider a superconductor with two impurities at positions $\vec{r}_1$ and $\vec{r}_2$:
\begin{equation}
H
=
\int d^2r\,
\Psi^\dagger_\alpha(\vec{r})
\left[
H^0_{\alpha\beta}(\vec{r})
+
U_{\alpha\beta}(\vec{r})
\right]
\Psi_\beta(\vec{r}),
\label{eq:hamiltonian}
\end{equation}
$\alpha,\beta=1,\ldots,4$. Here the band Hamiltonian of the superconductor in the absence of disorder is taken to be $H^0=\frac{1}{2}\hbar^2 \vec k^2$, and
$U_{\alpha\beta}(\vec{r})
=
\sum_{j=1,2}
U_j
(\tau_z\sigma_0)_{\alpha\beta}
\delta(\vec{r}-\vec{r}_j)
$
is disorder potential describing two nonmagnetic impurities, where the $\tau$ and $\sigma$ matrices act on particle-hole and spin degrees of freedom, respectively. In the basis
$
\Psi_{\vec{k}}
=
(c_{\vec{k},\uparrow},
c_{\vec{k},\downarrow},
c^\dagger_{-\vec{k},\uparrow},
c^\dagger_{-\vec{k},\downarrow})^T
$, the Bogoliubov-de Gennes Hamiltonian $H^0$ reads
\begin{equation}
H^0(\vec{k})
=
\begin{pmatrix}
\xi_{\vec{k}} & \Delta(\vec{k})(i\sigma_y)\\
\Delta^*(\vec{k})(-i\sigma_y) & -\xi_{\vec{k}}
\end{pmatrix},
\label{eq:bdg}
\end{equation}
yielding the Green's functions
\begin{equation}
G^{(0)}(i\omega_m,\vec{r},\vec{r}')
=
\int \frac{d^2k}{(2\pi)^2}
\frac{
e^{i\vec{k}\cdot(\vec{r}-\vec{r}')}}
{i\omega_m-H^0(\vec{k})}.
\label{eq:g0}
\end{equation}

\begin{figure}
\centering
\includegraphics[width=\linewidth]{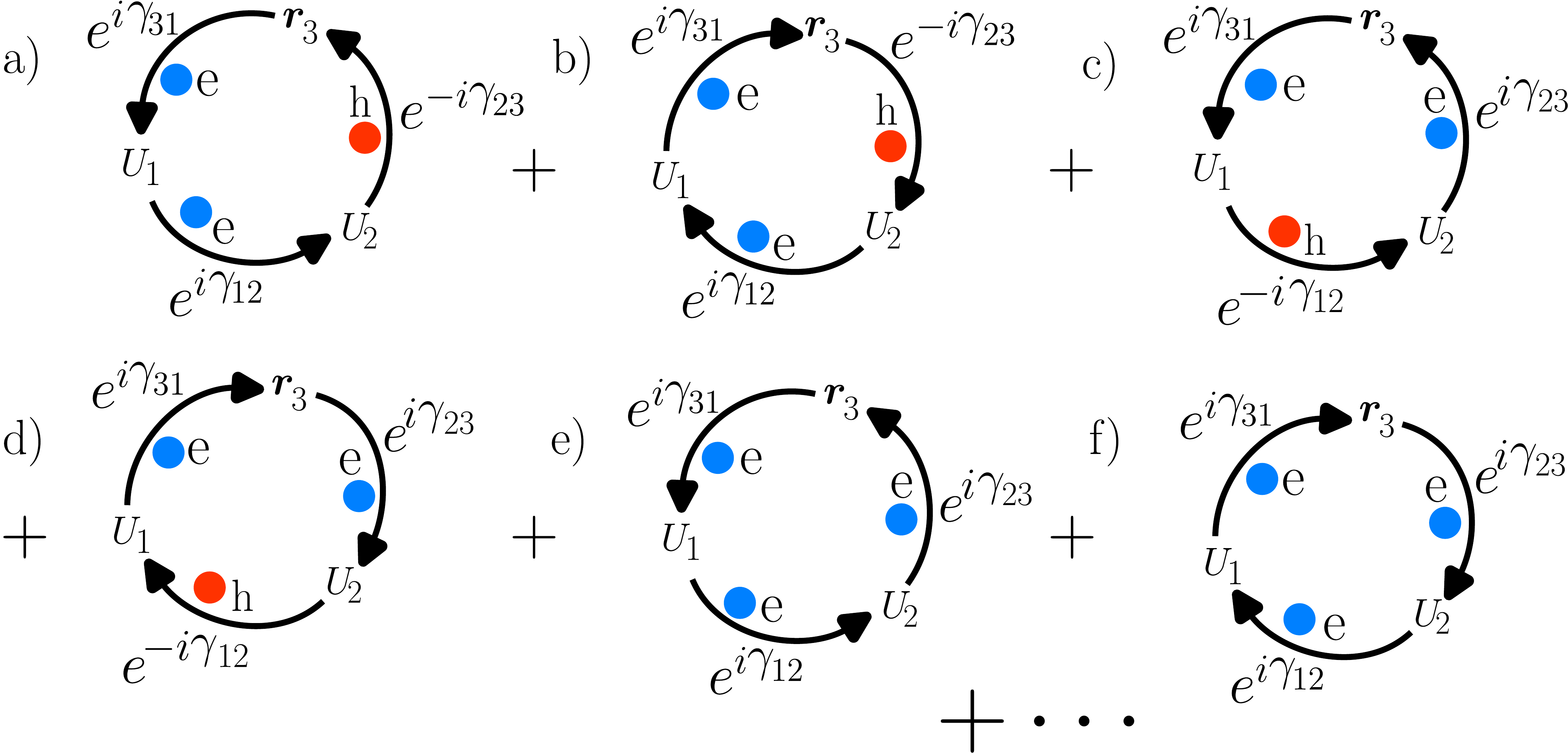}
\caption{
Diagrammatic contributions to TDOS responsible for QPI effects at second order in impurity potential (Eqs.~\eqref{eq:tseries} and \eqref{eq:delta_g2}). The quasiparticle (qp) and quasihole (qh) contributions $g_+$ and $g_-$ in Eq.~\eqref{eq:eilenberger} are shown separately: (a and b) qp-qh and (c--f) qp-qp contributions; qh-qh contributions are analogous to qp-qp contributions. Only the diagrams in (a and b) produce the QPI fringes of interest. The diagrams (a--d) vanish in a normal metal.
}
\label{fig:fig2}
\end{figure}

In the presence of disorder scattering, Green's function can be written as a perturbative series in terms of unperturbed Green's functions \cite{ref57,ref58,ref59},
\begin{align}
G(i\omega_m,\vec{r},\vec{r}')
&=
G^{(0)}_{\vec{r}-\vec{r}'}
+
\sum_j
G^{(0)}_{\vec{r}-\vec{r}_j}
T_j
G^{(0)}_{\vec{r}_j-\vec{r}'}
\nonumber\\
&\quad+
\sum_{j\neq j'}
G^{(0)}_{\vec{r}-\vec{r}_j}
T_j
G^{(0)}_{\vec{r}_j-\vec{r}_{j'}}
T_{j'}
G^{(0)}_{\vec{r}_{j'}-\vec{r}'}
+\ldots,
\label{eq:tseries}
\end{align}
where, for conciseness, we suppressed the dependence on discrete frequencies $i\omega_m$ in $T$'s and $G$'s. Here the $T$-matrices capture interaction with individual impurities renormalized by multiple scattering processes,
\begin{equation}
T_j(i\omega_m)
=
\left[
\hat I
-
(U_j\tau_z\sigma_0)
G^{(0)}_{\vec{r}=0}(i\omega_m)
\right]^{-1}
U_j\tau_z\sigma_0 .
\label{eq:tmatrix}
\end{equation}
Below we use these quantities to evaluate the tunneling conductance in coordinate space (Eq.~\eqref{eq:tdos}).

The microscopic mechanism by which the superconducting gap function becomes imprinted on particle-hole interference patterns is most transparent at distances greater than $\lambda_F$, the Eilenberger limit \cite{ref60}. In this case, the Green's function for a superconductor with gap $\Delta(\vec{k})$ simplifies considerably\footnote{See SI Appendix for derivation of Green’s functions and tunneling density of states in the presence of disorder scattering in the Eilenberger regime.}\footnote{The Eilenberger Green’s function is only valid in the $r_{ij} \gg \lambda_F$ limit, and therefore, the
$1/\sqrt{r_{ij}}$ divergence for small $r_{ij}$ is not a physical effect. Hence, we regularized $1/\sqrt{r_{ij}} \rightarrow 1/(r_{ij}^2 + 3 \lambda_F^2)^{1/4}$ while generating the plots.}, giving a sum of quasiparticle and quasihole contributions:
\begin{align}
G^{(0)}&(i\omega_m,\vec{r})
=
e^{i\gamma_r} g_+(\vec{r})
+
e^{-i\gamma_r} g_-(\vec{r}),
\nonumber\\
g_+(\vec{r})
&=
A_{\vec r}\tau_0+B_{\vec r}\tau_z
+
C_{\vec r}\tau_+(i\sigma_y)
+
C_{\vec r}^\dagger\tau_-(-i\sigma_y),
\nonumber\\
g_-(\vec{r})
&=
A_{-\vec{r}}\tau_0
-
B_{-\vec{r}}\tau_z
+
C_{-\vec{r}}\tau_+(i\sigma_y)
+
C^\dagger_{-\vec{r}}\tau_-(-i\sigma_y),
\label{eq:eilenberger}
\end{align}
where we denote
$
(A_{\vec r},B_{\vec r},C_{\vec r})
=
\frac{D_{\vec r}}{\Omega_{m,\vec r}}
\left(
i\omega_m,
i\Omega_{m,\vec r},
\Delta(\vec{k}_r)
\right),
$
$
\Omega_{m,\vec r}=\sqrt{\omega_m^2+|\Delta(\vec{k}_{\vec r})|^2}$,
$\tau_\pm=\frac{\tau_x\pm i\tau_y}{2}$,
and
$D_{\vec r}=-\sqrt{\frac{k_F}{8\pi v_F^2 r}} e^{-r\Omega_{m,\vec r}/v_F}$.
Here $\vec{r}$ denotes $\vec{r}-\vec{r}'$, and $\gamma_r=k_Fr-\pi/4$ is distance in $k_F^{-1}$ units, with $\gamma_r\gg1$ in our case. $\Delta(\vec{k}_r)$ denotes $\Delta(\vec{k})$ with $\vec{k}$ collinear with $\vec{r}$ and $|\vec{k}|=k_F$. For a spin-polarized triplet superconductor, we have to replace the $\pm i\sigma_y$ terms with unity. Here, $g_\pm$ denote particle-like and hole-like contributions to the propagator, where opposite phase factors $e^{\pm i\gamma_r}$ reflect counterpropagation of electrons and holes.

The particle-hole interference arises from the expansion in $T$-matrices $T_j$, $j=1,2$, at second order (the last term in Eq.~\eqref{eq:tseries}), where the quasiparticle in one side of the triangle diagram is hole-like ($g_{-}$) and another side is particle-like ($g_{+}$) (Fig.~\ref{fig:fig2}), i.e., the resultant contributions are terms like
$
{\rm tr}
\left[
g_+(\vec{r}_{31})
\tau_3
G^{(0)}_{\vec{r}_{12}}
\tau_3
g_-(\vec{r}_{23})
\right]
$ 
and permutations thereof. In the standard framework, calculating the tunneling density of states from BdG Green's functions involves projecting on the electronic component as ${\rm tr} [{\rm Im}(P_{ee}\delta G)]$, where $P_{ee}=(1+\tau_3)/2$. However, owing to the presence of particle-hole symmetry, the projector $P_{ee}$ can be dropped without changing the answer (see SI Appendix, section B). For two impurities of strength $U_0$ at points $\vec{r}_1$ and $\vec{r}_2$ in a spin-singlet superconductor,
\begin{align}\label{eq:delta_g2}
	&
	\frac{1}{2}{\rm{tr}} \delta G^{(2)}(i \omega_m,\vec r_3)
	\!=\!
	\left(\frac{2k_F}{\pi v_F^2}\right)^{3/2} \frac{e^{-\left(\sum_{i>j} r_{ij} \Omega_{ij}\right)/v_F}}{\sqrt{r_{12} r_{23} r_{31}}} 
\nonumber \\
&
\times \frac{i\omega_m U_0^2}{\Omega_{12}\Omega_{23}\Omega_{31}} 
\Bigg[\sin\gamma_{12}  \sin(\gamma_{23} - \gamma_{31}) \Omega_{12}(\Omega_{23}-\Omega_{31}) \nonumber \\
&  \qquad + \cos\gamma_{12} \cos(\gamma_{23}-\gamma_{31}) 
\Big( \Omega_{23} \Omega_{31} -\omega_m^2 
\\
&+{\rm{Re}}[\Delta_{23} \bar\Delta_{31} - \bar\Delta_{12}(\Delta_{23}+\Delta_{31})]\Big)
\Bigg] +\ldots,
\nonumber 
\end{align}
where $r_{ij}=|\vec{r}_i-\vec{r}_j|$, $\Delta_{ij}=\Delta(\vec{k}_{\vec{r}_{ij}})$, and $\gamma_{ij}=k_F r_{ij}-\pi/4$. For conciseness, the subscript $m$ in $\Omega_{ij}$ has been dropped. The terms proportional to $\cos(\gamma_{23}-\gamma_{31})$ and $\sin(\gamma_{23}-\gamma_{31})$ give rise to Young's interference fringes, whereas other terms---marked by dots---proportional to $\cos(\gamma_{23}+\gamma_{31})$ and $\sin(\gamma_{23}+\gamma_{31})$ do not produce such fringes. 
These terms correspond to diagrams in Fig.~\ref{fig:fig2} (a and b) and (c--f), respectively.
The visibility of qp fringes is proportional to
$
{\rm Re}\left(
|\Delta_{\vec{r}_3}|^2
-
\Delta_{\vec{r}_3}\Delta^*_{\vec{r}_{12}}
\right)$
at large distances, since in this limit
$
\Omega_{23}\Omega_{31}-\omega_m^2
\approx
|\Delta_{\vec{r}_3}|^2
\approx
\Delta_{23}\Delta^*_{31},
$
giving Eq.~\eqref{eq:visibility}.

The tunneling conductance $dI/dV$ measured at position $\vec{r}$ at zero bias can be expressed in a convenient form as (see ref.~\cite{ref61}, and also, SI Appendix, section B),
\begin{equation}
\left.
\frac{dI}{dV}
\right|_{V=0}
(\vec{r})
=
4e^2 |t_{\rm STM}|^2\nu_0 T
\sum_{\omega_m>0}
{\rm tr}\,
{\rm Re}
\left[
\frac{dG(z=i\omega_m,\vec{r},\vec{r})}{dz}
\right],
\label{eq:tdos}
\end{equation}
where $t_{\rm STM}$ is the STM tip-to-superconductor tunneling amplitude and $\nu_0$ is the density of states at the Fermi level. This quantity is nonzero at $T>0$ even for a gapped superconductor. The two impurity contribution to this quantity, plotted in Figs.~\ref{fig:fig1} and ~\ref{fig:fig4} displays radial fringes of Young's interference modulated with nodal lines. While Young's interference is generic for particle-hole interference near two impurities, the pattern of nodal lines is unique for the superconducting order type, as discussed above.

These results can be used to predict how the visibility of fringes scales with $\Delta$. In the $T$-matrix expansion, the second-order terms are smaller than the first-order terms by a relative factor of $U_0/(v_F\hbar\sqrt{\lambda_F r_{12}})$. The typical magnitude of the interference term in TDOS is therefore
$
4e^2|t_{\rm STM}|^2\nu_0 U_0^2
\left(
\frac{k_F}{2\pi\lambda_F v_F^2}
\right)^{3/2}
h(T,\Delta),
$
where $h(T,\Delta)$ behaves as $T^3/\Delta^3$ for $T\ll\Delta$ and as $\Delta^2/T^2$ for $\Delta\lesssim T\le T_c$.

While qp interference is a generic property of superconductors with unconventional pairing (such as topological $d+id$ or nontopological $d_{x^2-y^2}$ superconductor), it is absent for an angle-independent $s$-wave gap function. In this case, only Friedel-like oscillations are expected. In our toy model, at second order in impurity potential,
\begin{equation}
{\rm tr}\,
\delta G^{(2)}(i\omega_m,\vec{r})
\sim
\frac{
U_0^2
\left(k_F/2\pi\right)^{3/2}
(i\omega_m)}
{\pi v_F^3\sqrt{\omega_m^2+\Delta^2}}
\frac{
\cos(\gamma_{12}+\gamma_{23}+\gamma_{31})}
{\sqrt{r_{12}r_{23}r_{31}}}.
\label{eq:swave_second_order}
\end{equation}
To understand this behavior, we note that in this case the matrices $g_{+/-}(\vec{r}_{ij})$ do not depend on the quasiparticle propagation direction $\hat{\vec{r}}$. Due to unitarity of the Green's function in the Eilenberger limit $r\gg\lambda_F$ \cite{ref60,ref62}, the matrix $g_+(\vec{r})\tau_3g_-(\vec{r})$ is identically zero for BCS $s$-wave superconductor, which ensures that the particle-hole contribution vanishes identically. This cancellation holds only when $\Delta(\vec{k})$ is angle independent.

Last, we discuss ways to enhance quasiparticle interference in the SOPT approach and extract nodal lines from experimental TDOS spatial maps around two impurities, in which they occur alongside Friedel oscillations. The separation of particle-hole and particle-particle contributions proves feasible because Friedel oscillations have wavenumbers $\approx 2k_F$, whereas the particle-hole contribution, being essentially Young's interference, occurs at markedly lower wavenumbers $k\lesssim1/|\vec{r}_1-\vec{r}_2|$. In SI Appendix, section F, we demonstrate that the different wavenumbers of these signals allow the Young’s interference pattern, along with its nodal beams, to be isolated with Fourier filtering. Fourier analysis is routinely used in FT-STS \cite{ref55} method involving many impurities to map out the Fermi surface. Here we propose a similar technique, but after removing Friedel oscillations in momentum space, the signal needs to be inverse transformed to real space. These characteristic wavenumbers are functions of the band structure and geometric configuration of the impurities, and their presence, as well as their nodal lines, do not depend on the energy bias at which they are probed, as long as the energy scale is not too large compared to $\Delta$.

\section{Application to Realistic Band Structures: Strontium Ruthenate}

Having established how the SOPT method can be utilized in an ideal scenario involving a single parabolic band with isotropic dispersion, we now demonstrate how it works in realistic band structures, and we consider Sr$_2$RuO$_4$ as an example. We choose Sr$_2$RuO$_4$ for a number of reasons. It is made of weakly coupled layers, which make the electron dynamics effectively two-dimensional, and it is well suited for performing STM experiments on the surface. Moreover, we find that the square nature of the bands in this system spatially separates the Friedel oscillations from the regions where two-impurity interference is observed, eliminating the need for Fourier filtering. Also, despite overwhelming evidence for unconventional pairing, long believed to be $p+ip$ but contested recently \cite{ref63,ref64}, the exact nature of pairing in this superconductor remains unresolved \cite{ref65,ref66,ref67}, which makes it a promising candidate for the SOPT technique. Sr$_2$RuO$_4$ has a cleavable perovskite structure with atomically clean surfaces, ideal for STM spectroscopy \cite{ref68,ref69}. Recent studies \cite{ref54} indicate that the pairing is likely either $d_{xz}+id_{yz}$ or $s'+id_{x^2-y^2}$. The $s'+id_{x^2-y^2}$ phase breaks time-reversal symmetry, but is not topological. Interestingly, in this phase, both $s'$ and $d_{x^2-y^2}$ components were predicted to have adjacent nodes (see ref.~\cite{ref54}). In this case, static `accidental' nodal beams may be observed even though TRS is broken, in departure from the general rule that TRS breaking prohibits static nodal beams (see table in Fig.\ref{fig:fig1}(c)).

\begin{figure}
\centering
\includegraphics[width=\linewidth]{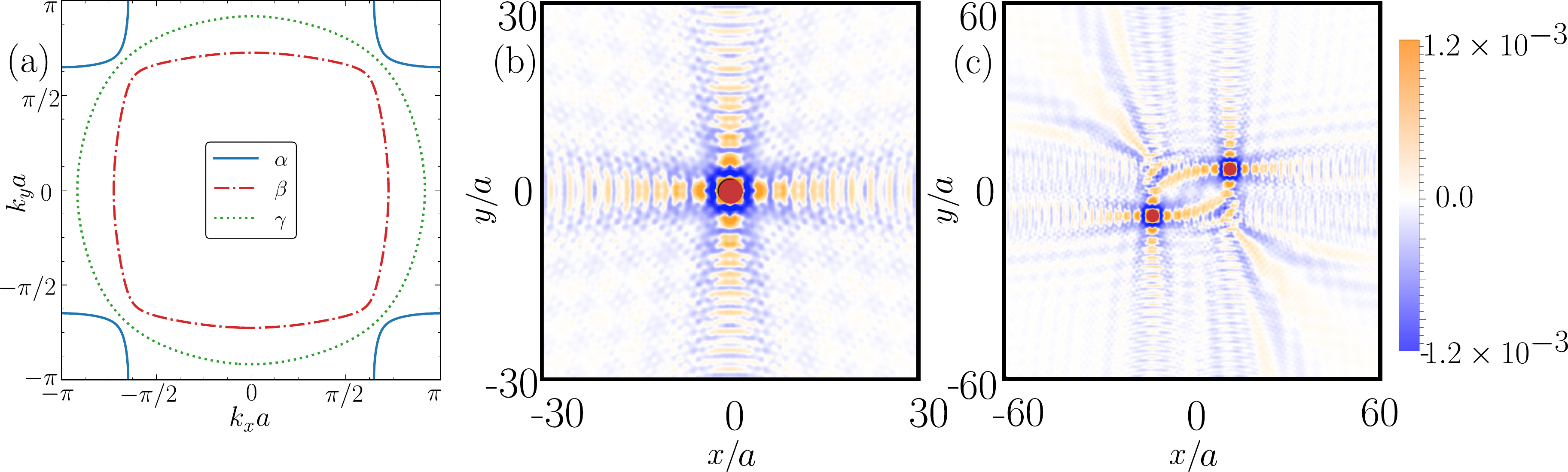}
\caption{
(a) Fermi surface of Strontium Ruthenate, obtained from the normal metal Hamiltonian. (b) Friedel oscillation in TDOS around an impurity in the $d+id$ superconducting phase due to the square-shaped $\alpha,\beta$ bands, including effects of Wannier orbitals. Here $a$ is lattice constant, and TDOS is expressed in units of $4e^2|t_{\rm STM}|^2\nu_0U_0/(t_1a^2)^2$, where $t_{\rm STM}$ and $U_0$ are STM tunneling coupling and the impurity potential strength defined in Eqs.~\eqref{eq:tdos} and \eqref{eq:tmatrix}. Realistic parameter values were used: $\Delta_0=t_1/100$, $k_BT=0.5\Delta_0$. (c) Joint plot of the first-order (Friedel oscillations) and the second-order contributions, demonstrating that they occur in separate directions in coordinate space. The second-order contribution is artificially enhanced 100 times for clarity. Only the second-order contribution is sensitive to the symmetry of the pairing, as discussed in Figs. \ref{fig:fig4} and \ref{fig:fig6}.
}
\label{fig:fig3}
\end{figure}

In real materials, the QPI pattern deviates from the toy model described above due to a variety of reasons. First, the Fermi surface is not perfectly circular, and the Fermi velocity and the Fermi momentum are not isotropic. Second, the variation of tunneling density of states on the sublattice level is modified by the anisotropic structure of Wannier functions. Moreover, the superconducting order parameter and spin-orbit couplings can cause band mixing, modifying the experimentally observed QPI.

Sr$_2$RuO$_4$ has three Fermi surface components, of which two (denoted $\alpha,\beta$) are square shaped, and the third one ($\gamma$), is nearly circular; see Fig.~\ref{fig:fig3}(a). We start with the BdG Hamiltonian in block form, where the diagonal part constitutes a $3\times3$ matrix describing the three bands \cite{ref70,ref71,ref72}, and the superconducting gap function is also described by a $3\times3$ matrix, as described in SI Appendix, section D. The spatial variation of tunneling conductance in the sublattice scale is found from the lattice Green's function (similar to Eq.~\eqref{eq:g0} but with coordinates $\vec{R},\vec{R}'$ on the crystal lattice) $G^{(0)}_{\mu\nu}$ and Wannier functions (equivalent to the BdG+W method \cite{ref73,ref74}, see SI Appendix, section C):
\begin{equation}
G^{(0)}(i\omega_n,\vec{r},\vec{r}')
=
\sum_{\mu,\nu}
\sum_{\vec{R},\vec{R}'}
G^{(0)}_{\mu\nu}
(i\omega_n,\vec{R},\vec{R}')
w_{\mu,\vec{R}}(\vec{r})
w^*_{\nu,\vec{R}'}(\vec{r}'),
\label{eq:wannier}
\end{equation}
where $w_{\mu,\vec{R}}(\vec{r})$ denotes the Wannier function of orbital $\mu$ centered at lattice point $\vec{R}$, evaluated at a continuum point $\vec{r}$. Among the multiple superconducting phases theoretically proposed for Sr$_2$RuO$_4$, we identify three interesting and actively debated phases, namely $d+id$, $p+ip$, and $d_{x^2-y^2}$, (and we also consider a hypothetical s-wave superconductor), to identify the utility of the tomography method in identifying their unique signatures.
We numerically compute the lattice Green's function from the BdG Hamiltonian, and plug in this Green's function into Eqs.~\eqref{eq:tseries} and \eqref{eq:tdos} to obtain local TDOS patterns for these order parameters. The effect of convoluting the lattice Green's function with Wannier orbitals is illustrated in Fig.~\ref{fig:fig5}.

\begin{figure}
\centering
\includegraphics[width=\linewidth]{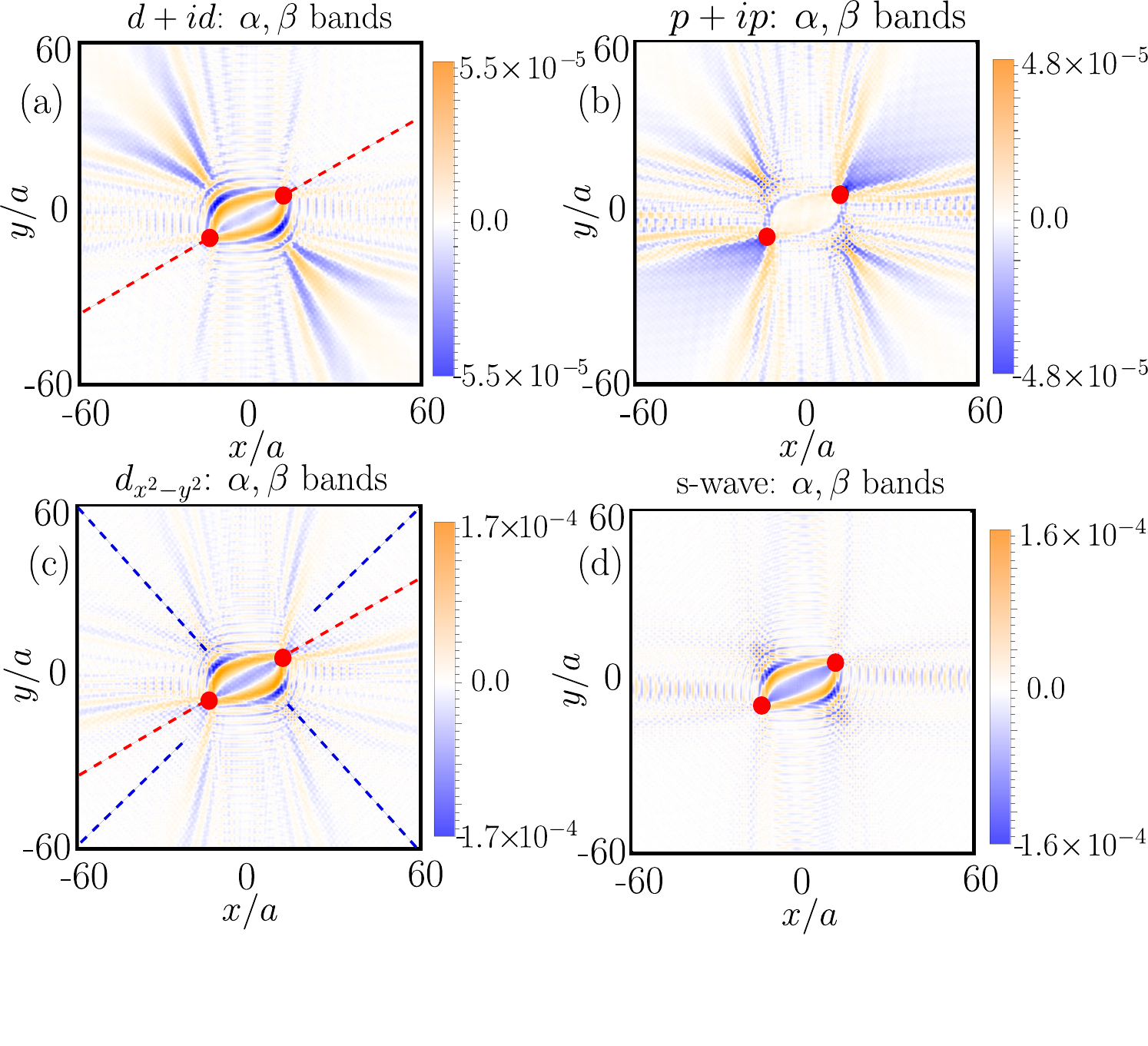}
\caption{
Second order contribution to $dI/dV$ for various order parameters in Sr$_2$RuO$_4$: The quantity $dI/dV$ obtained for the $\alpha$ and $\beta$ bands, including the effect of Wannier orbitals, for orders (a) $d+id$, (b) $p+ip$, (c) $d_{x^2-y^2}$, and (d) $s$-wave superconductor. Young’s interference appears when the gap function has angular dependence, whether topological or nontopological (a, b, c), and is absent for an angle-independent s-wave gap function (d). TDOS is plotted in the units of $4e^2|t_{\rm STM}|^2\nu_0U_0^2/(t_1a^2)^3$, for two impurities a distance $30a$ apart rotated by angle $30^\circ$ with respect to the $x$ axis. The spin-triplet $p+ip$ superconductor (b) has a distinct feature that there is a strong second-order interference effect on the line joining the impurities when the impurities are not aligned with the axes of the lattice. We have set the strength of the gap strength $\Delta_0=t_1/100$ (details in SI Appendix, section D) and temperature $k_BT=0.5\Delta_0$.
}
\label{fig:fig4}
\end{figure}

\begin{figure}
\centering
\includegraphics[width=\linewidth]{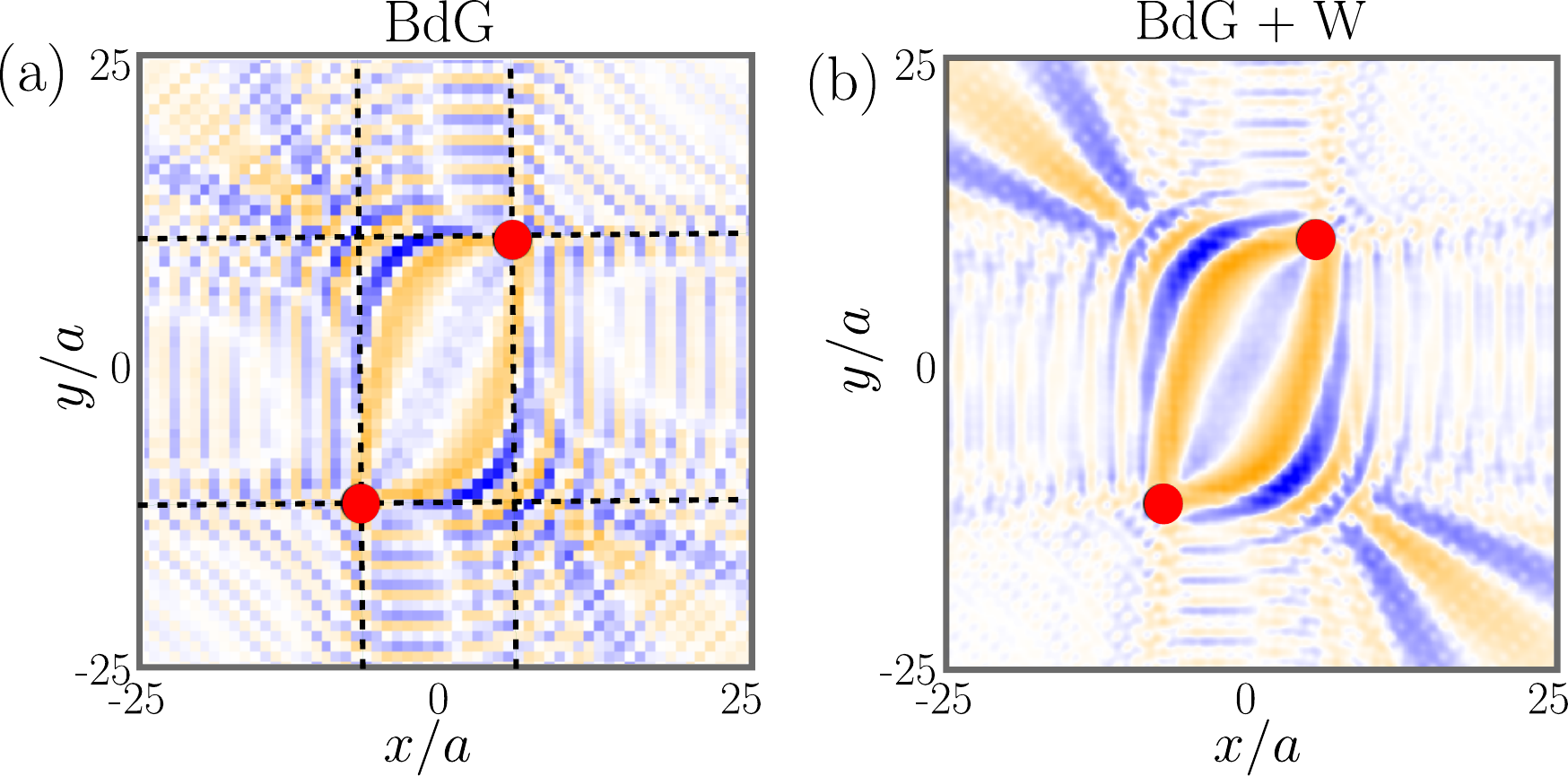}
\caption{
Comparison between second-order contribution of the $\alpha$ and $\beta$ bands to TDOS in the (a) BdG and (b) BdG+W methods (see Eq.~\eqref{eq:wannier},  refs.~\cite{ref73,ref74}). In the latter, the Green's functions are convoluted with the Wannier functions, resulting in a smooth curve on the sublattice lengthscale. Here we have plotted the TDOS of the $d+id$ phase of Strontium Ruthenate near two impurities aligned at $60^\circ$ with the $x$ axis. The distance between the two impurities is $30a$. Notice how the Young's interference fringes become noticeable when the effect of the Wannier orbitals are included. In these figures, orange indicates a positive sign of signal, and blue indicates negative sign, while white indicates zero. In panel (a) the dotted black lines show the expected position of the Friedel oscillations (depicted in Fig.~\ref{fig:fig3}(b)). Notice that the hyperbolic patterns emerge where the two horizontal and vertical lines intersect, a pattern that is also observed in Fig.~\ref{fig:fig6}. The TDOS is plotted in the same units as in Fig.~\ref{fig:fig4}.
}
\label{fig:fig5}
\end{figure}

We compute the total second-order contribution to TDOS, and for the nearly circular $\gamma$ band, we find all the general features of the SOPT method discussed above: (i) the $s$-wave state does not feature Young's interference fringes; (ii) The $d_{x^2-y^2}$ superconductor hosts static nodal beams along the nodes of the superconductor, at $45^\circ$ and $135^\circ$; and (iii) The singlet topological superconductors with inversion symmetry (e.g. $d+id$), retain a rotating nodal beam along the line joining the impurities (as the quasiparticle propagation along $\pm\vec{k}$ produces the same effect), whereas spin-triplet superconductors (e.g. $p+ip$) do not host such a rotating nodal beam. This effect demonstrates how our previous results are readily applicable in realistic materials with nearly circular bands.

In comparison, the square-like $\alpha,\beta$ bands have a unique QPI imprint (Fig.~\ref{fig:fig4}), as the quasiparticles are mostly constrained to move along the $\hat{x},\hat{y}$ directions. We find that for exactly square-shaped bands the two-impurity contribution to TDOS vanishes entirely unless the two impurities are aligned with the sides of the square, i.e. along either $\hat{x}$ or $\hat{y}$. However, since these bands acquire some nonsquareness due to band mixing, the realistic bands do produce interference effects at the second order, but the resulting pattern is qualitatively different from traditional Young's interference patterns. We find that hyperbolic fringes are dominantly pronounced in the directions which are not aligned with the $x,y$ axes with the square bands, and hence they are better pronounced when the impurities are not aligned with the crystal axes. Each superconducting order parameter candidate produces a unique pattern. We find that (i) an $s$-wave superconductor with uniform gap does not produce Young's interference fringes, demonstrating that our previous claim for an ideal parabolic band still holds in realistic band structures; (ii) As illustrated with a toy model, the realistic model of a $d_{x^2-y^2}$ superconductor does not feature Young's interference fringes along directions where it becomes gapless, i.e. it hosts a static nodal beam; and (iii) the topological superconductors with and without inversion symmetry, namely $d+id$ and $p+ip$, can be distinguished by the presence of the rotating nodal beam along the line joining the impurities.

\begin{figure}
\centering
\includegraphics[width=\linewidth]{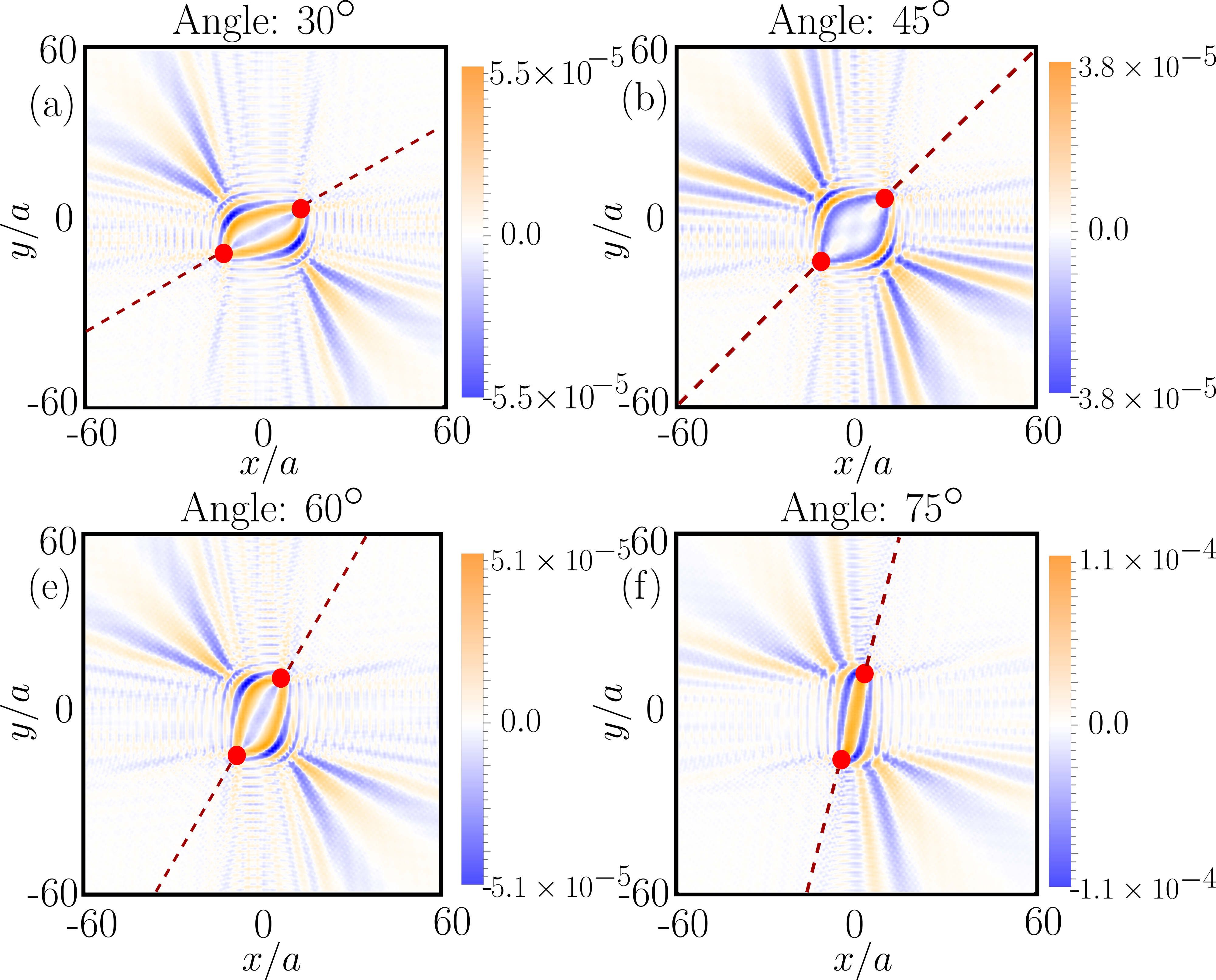}
\caption{
Signatures of a $d+id$ phase in the two-impurity QPI tomography for the $\alpha,\beta$ bands of Strontium Ruthenate. Young's interference signals an angle-dependent pairing gap $\Delta(\vec k)$, topological or nontopological. The second-order contribution to TDOS is shown for impurities' rotation angles (a) $30^\circ$, (b) $45^\circ$, (c) $60^\circ$, and (d) $75^\circ$ with respect to the $x$ axis. Notably, Young's interference fringes originate from points where the horizontal and vertical lines drawn centered at the impurities intersect. The intersection point moves as the impurities are rotated, enhancing the visibility of QPI fringes. The rotating nodal beams are depicted as dashed red lines. The TDOS is plotted in the same units as in Fig.~\ref{fig:fig4}.
}
\label{fig:fig6}
\end{figure}

Now we discuss the behavior of TDOS of the $\alpha$ and $\beta$ bands (the ones which couple strongly with STM) in Strontium Ruthenate under controlled rotation of impurities. While static and rotating beams behave similarly to circular bands, the hyperbolas do not rotate uniformly with impurity rotation due to the highly nonisotropic nature of the bands, a feature not found in optical Young's interference experiments. Since the majority of quasiparticles on these square-shaped bands have a velocity along $\hat{x}$ and $\hat{y}$, the particle-hole interference is very different from a circular band. Unlike a circular band case, the hyperbolic fringes now emerge from a point of intersection of fictitious horizontal and vertical lines drawn from the two impurities. Here, we demonstrate this behavior for the $d+id$ order parameter in Fig.~\ref{fig:fig6}. Since the Young's interference patterns along $x$ and $y$ axes are blocked due to the square nature of the bands, these hyperbolic patterns can only show up in the resulting four ``quadrants.'' The signal is highly diminished in the two quadrants that host the line joining the impurities due to the rotating nodal beam, and we find strong Young's interference signature in the other two quadrants. The patterns of rotating nodal beams are unique to each superconductor order parameter candidate. Therefore, with this method, different order parameter candidates in Sr$_2$RuO$_4$ can be distinguished.

Before concluding, we discuss the details of the realistic model and justifications of the approximations involved. In our calculation, we modeled impurities as point-like, nonmagnetic scatterers with isotropic, weak delta-function potential. However, real impurity potentials have a finite range and they feature anisotropic scattering. Our results remain valid for impurity potentials with a finite range so long as distances between the impurities are much larger than the Fermi wavelength. While the anisotropy can enhance scattering in some directions, quasiparticle interference fringes and their nodal structures should persist, with overall intensity modulated by anisotropy of the impurity potential. Here, we did calculations under the assumption of ``weak impurity strength'', for which the impurity potential is treated perturbatively up to second order and does not produce in-gap states. However, the ellipse and hyperbola-like patterns will prevail up to any order of perturbation theory and therefore remain valid for stronger impurities as long as the $T$-matrix theory is applicable. Also, qualitatively similar results are obtained when the Fermi sea is not centered at the $\Gamma$ point, and our results should remain valid even in graphene based superconductors where the Fermi sea is centered at $K$ or $K'$ points. Also, while the appearance of circular nodal rings in our calculations is a consequence of the circular Fermi surface and isotropic gap function magnitude utilized in our model, deviation from isotropy is expected to distort the shape of the nodal ring, but it would not disappear altogether. We expect qualitatively similar features in QPI in the presence of magnetic impurities.

\section{Conclusions}

The overarching goal of this work is to demonstrate feasibility of a real-space probe of superconducting order parameter $\Delta(\vec{k})$ based on the two-impurity quasiparticle interference (QPI) measured by spatially resolved STM probes. Unlike traditional STM experiments, which are essentially ``blind'' to the order parameter phase, this approach utilizes controlled rotation of two impurities to extract both the nodes of the order parameter and its phase winding. Despite being a well-established phenomenon, QPI has not been utilized as a diagnostic of gap-function $\vec{k}$-dependence, particularly phase winding.

A central result of this work is that spatial interference patterns near two impurities provide clear telltale signatures of $\Delta(\vec{k})$ angular dependence and its pairing symmetry. Nodes in the gap function correspond to static nodal beams, and the information about complex phase winding can be revealed by the rotating nodal beams of the Young's interference pattern near two impurities. Based on these findings, full reconstruction of the momentum dependence of $\Delta(\vec{k})$ from this technique appears feasible, and may serve as an interesting topic for future research.

In the recent years, a number of materials have emerged as candidate unconventional superconductors---some more controversially than others---including UPt$_3$, Sr$_2$RuO$_4$, URu$_2$Si$_2$, UTe$_2$, LaPt$_3$ (see refs. \cite{ref75,ref76} for a broader survey), and, recently, rhombohedral and moir\'e graphene and twisted bilayer MoTe$_2$ \cite{ref77,ref78,ref79,ref80}. The SOPT effect in the two-impurity interference is well suited for probing unconventional superconductors with layered structures and clean surfaces, allowing to use state-of-the-art STM spectroscopy to reveal the nature of pairing.

As an illustration of an application to realistic band structures, we consider Sr$_2$RuO$_4$, a system that has a number of distinct advantages for this study. The two-impurity interferometry is found to be capable of distinguishing different theoretically proposed competing order parameters for this enigmatic superconductor which has been a subject of ongoing debate. This therefore opens a realistic path toward resolving the superconducting order parameter in Sr$_2$RuO$_4$. More broadly, this work establishes two-impurity interferometry as a practical and appealing route to phase-sensitive superconducting gap tomography in real materials.

\section{Materials and Methods}

The contribution of quasiparticle interference to TDOS is evaluated using a BdG Hamiltonian describing quantum-coherent particle-hole scattering by two impurities. Scattering by impurities is described by Green's functions in coordinate space evaluated by the $T$-matrix power series expansion. The resulting TDOS displays two-impurity quasiparticle interference patterns with several distinct features: radial interference fringes and nodal beams, static and rotating. The behavior of the fringes and beams under controlled rotation of two impurities is used to infer the angle-dependent superconducting gap function and the order-parameter symmetry. This method is illustrated with a two-dimensional parabolic-band toy model and a two-dimensional model of Strontium Ruthenate. In SI Appendix we provide i) detailed derivations of the Eilenberger Green’s functions of the toy models of topological superconductors, ii) derivation of the tunneling conductance formula in Eq.~\eqref{eq:tdos}, iii) details of the Wannier orbitals, iv) details of the band Hamiltonian of Sr$_{2}$RuO$_4$, v) results for TDOS of $\gamma$ bands which do not strongly couple with STM, and vi) Fourier filtering of Young’s interference fringes.

\section*{Data, Materials, and Software Availability}

Code and data have been deposited in GitHub \cite{ref82}.

\begin{acknowledgments}
We thank Andrey Chubukov, Zhiyu Dong, Peter Hirschfeld, Steven Kivelson, Patrick Lee, Khachatur Nazaryan, Boris Spivak, and Eli Zeldov for useful discussions. This work was performed in part at the Aspen Center for Physics, which is supported by National Science Foundation grant PHY-2210452. It was also supported in part by the U.S.-Israel Binational Science Foundation (BSF).
\end{acknowledgments}

\end{document}


\title{Supplementary Information: Tomographic imaging of superconducting order using particle-hole interference} 



\author{Archisman Panigrahi, Vladislav Poliakov
and Leonid Levitov\\
\textit{\small Department of Physics, Massachusetts Institute of Technology, 77 Massachusetts Avenue, MA 02139}}

\maketitle
\onecolumngrid
	\appendix
	\numberwithin{equation}{section}
\section{Green's functions in the Eilenberger limit, $r \gg \lambda_F$}
	This section derives the semiclassical ($r \gg \lambda_F$) Green's functions for a toy-model superconductor with a 2D circular parabolic band and an arbitrary angle dependence of the gap function $\Delta(\vec k)$. The cases considered explicitly are a $d+id$ topological superconductor and a $d_{x^2-y^2}$ superconductor, whose local tunneling density of states (TDOS) patterns are shown in the main text. Although the Eilenberger limit is asymptotically valid for $r\gg \lambda_F$, it also captures the qualitative spatial structure of two-impurity quasiparticle interference (QPI) near the impurities. 
	
	QPI in real materials is more accurately described using lattice Green's functions. These Green's functions are introduced in Appendix C, and their use is illustrated in the main text for QPI in a multiband model of strontium ruthenate. It is nevertheless useful to begin with a single-band toy model with a circular Fermi surface, for which the Green's function can be evaluated in the semiclassical limit as described below.
	
	For a general BdG Hamiltonian, as given in Eq. 6 of the main text, the position-space Green's function is obtained by Fourier transforming $1/(i\omega_m - H^0(\vec k))$ and evaluating the integral in polar coordinates, setting $\vec r' = 0$:
	\be
	G^{(0)}(i\omega_m,\vec r) = \int \frac{d^2 \vec k }{(2\pi)^2} \frac{e^{i \vec k \cdot \vec r}}{i\omega_m - H^0(\vec k)}
	= \int_0^\infty \frac{k dk}{(2\pi)^2} \oint d\theta \frac{e^{i k r \cos (\theta-\theta_{\vec r})}}{i\omega_m - H^0(\vec k)}	
	,
	\ee
	where $\theta_{\vec r}$ is the angle between $\vec r$ and the $x$-axis. In the limit of large $r$, the exponential factor $e^{i k r \cos(\theta-\theta_{\vec r})}$ in the integral will rapidly oscillate except near the extrema of $\cos(\theta-\theta_{\vec r})$. 
    
    This behavior is captured by the complex saddle-point approximation. According to  this method, the largest contribution to the integrals of the form 
    \[
    \int_{0}^{2\pi} e^{i b \cos(\theta-\theta_0)} h(\theta) d\theta,
    \]
    for a smooth, periodic function $h(\theta)$, in the limit of $b \gg 1$, comes from the points of extrema of the rapidly oscillating exponent of the exponential function, which occur at $\theta = \theta_0$, $\theta = \theta_0 + \pi$. 
	As a result, the leading contribution to  $G^{(0)}(i\omega_m,\vec r)$ arises from $\vec k$ parallel and antiparallel to $\vec r$, describing Bogoliubov quasiparticles---superpositions of electrons and holes---propagating along $\vec r$ and $-\vec r$. 
	
	Accordingly, the notation
	$ H^0(k \vec{\hat{r}})$ denotes the $k$-space BdG Hamiltonian with momentum $\vec k$ pointing along $\vec{\hat{r}}$. In the main text, the abbreviation $\vec k_{\vec r}$ denotes this quantity. Similarly, $\Delta(\theta_{\vec r})$ denotes $\Delta(\vec k)$ for $\vec k$ aligned with $\vec{\hat{r}}$ and with magnitude $k_F$, since pairing is significant only near the Fermi surface. Likewise, $\Delta(\theta_{\vec r} + \pi)$ is identical to $\Delta(\theta_{-\vec r})$.
	
	Applying the saddle-point method gives
	\begin{equation}
		\int_0^{2\pi} e^{i b \cos(\theta-\theta_0)} h(\theta) d\theta \approx \sqrt{\frac{2\pi}{b}} \left[h(\theta_0) e^{i(b-\frac{\pi}{4})} + h(\theta_0 + \pi) e^{-i(b-\frac{\pi}{4})}\right] +O\left(\frac{1}{b^{3/2}}\right)
		,
	\end{equation}
	where $h(\theta)$ is an arbitrary smooth function. 
	Using this result for the angular integration gives the large-distance behavior of the Green's function: 
	\be
	G^{(0)}(i\omega_m,\vec r)
	\approx  \int_0^\infty \frac{k dk}{(2\pi)^2} \sqrt{\frac{2\pi}{k r}} \left[\frac{e^{i (k r - \pi/4)}}{i\omega_m - H^0(k \vec{\hat{r}})} + \frac{e^{-i (k r - \pi/4)}}{i\omega_m - H^0(-k \vec{\hat{r}})}\right]
	\ee
	The dominant contribution to this integral is expected to come from the region $\delta k\approx \sqrt{\omega_m^2+\Delta^2}/v_F\ll k_F$ near the Fermi level. Therefore,
	\be
	G^{(0)}(i\omega_m,\vec r)
	\approx  \int_0^\infty  \frac{k_F dk}{(2\pi)^2} \sqrt{\frac{2\pi}{k_F r}} \left[\frac{e^{i (k r - \pi/4)}}{i\omega_m - H^0(k \vec{\hat{r}})} + \frac{e^{-i (k r - \pi/4)}}{i\omega_m - H^0(-k \vec{\hat{r}})}\right]
	\ee
	Next, the dispersion is linearized near the Fermi surface using $\xi = v_F (k-k_F)$ as the integration variable.
	Since the dominant contribution comes from states near the Fermi level, the limits of integration over $\xi$ may be extended to $-\infty<\xi<\infty$, giving
	\begin{equation}
		\begin{aligned}
			G^{(0)}(i\omega_m, \vec r) &\approx -\sqrt{\frac{k_F}{(2\pi)^3 v_F^2 r}} \int_{-\infty}^{\infty} \frac{d\xi}{v_F} \left[\frac{e^{i (k_F r - \pi/4)} e^{i\xi r/v_F}}{i\omega_m - \left(\xi \tau_z + 
				\Delta(k \vec{\hat{r}})\tau_+ (i\sigma_y) + {\Delta^*}(k \vec{\hat{r}})\tau_- (-i\sigma_y)\right)} \right. \\ 
			&\qquad\qquad\qquad\qquad\qquad\qquad + \left. \frac{e^{-i (k_F r - \pi/4)}e^{-i\xi r/v_F}}{i\omega_m - \left(\xi \tau_z + 
				\Delta(-k \vec{\hat{r}})\tau_+ (i\sigma_y) + {\Delta^*}(-k \vec{\hat{r}})\tau_- (-i\sigma_y)\right)}\right]
		\end{aligned}
	\end{equation}
	To perform the integration over $\xi$, the Green's function matrix is rationalized, yielding
	\begin{equation}\label{app:eq:after-multiplying-num-den}
		\begin{aligned}
			G^{(0)}(i\omega_m, \vec r) &= -\sqrt{\frac{k_F}{(2\pi)^3 v_F^2 r}} \int_{-\infty}^{\infty} d\xi  \left[\frac{e^{i \gamma_r} e^{i\xi r/v_F} [i\omega_m + \left(\xi \tau_z + 
				\Delta(k \vec{\hat{r}})\tau_+ (i\sigma_y) + {\Delta^*}(k \vec{\hat{r}})\tau_- (-i\sigma_y)\right)]}{\xi^2 + \omega_m^2 + |\Delta(k \vec{\hat{r}})|^2} \right. \\
			&\qquad\qquad\qquad\qquad\qquad\qquad+ \left. \frac{e^{-i \gamma_r}e^{-i\xi r/v_F} [i\omega_m + \left(\xi \tau_z + 
				\Delta(-k \vec{\hat{r}})\tau_+ (i\sigma_y) + {\Delta^*}(-k \vec{\hat{r}})\tau_- (-i\sigma_y)\right)]}{\xi^2 + \omega_m^2 + |\Delta(-k \vec{\hat{r}})|^2}\right]
		\end{aligned}
	\end{equation}
	Integrals over $\xi$ can be evaluated using the identities 
	\begin{equation}
			\int_{-\infty}^{\infty}  \frac{e^{i a x}}{x^2 + b^2} dx = \frac{\pi}{|b|} e^{-| a b|}, \qquad 
			\int_{-\infty}^{\infty} \frac{e^{i a x} x}{x^2 + b^2} dx = i\pi \sgn(a) e^{-|a b|},
	\end{equation}
	derived by contour integration for real parameters $a, b$. 
	Applied to the integrals in Eq.\eqref{app:eq:after-multiplying-num-den}, 
	it yields
	%
	\begin{equation}\label{eq:G_general}
		\begin{aligned}
			G^{(0)}(i\omega_m,\vec r)
			& = E e^{i\gamma_r} \left[e^{-\frac{r}{v_F}\sqrt{\omega_m^2 + |\Delta(k \vec{\hat{r}})|^2}} \left(\left(i\omega_m + \Delta(k \vec{\hat{r}})\tau_+ (i\sigma_y) + {\Delta^*}(k \vec{\hat{r}})\tau_- (-i\sigma_y)\right)\frac{\pi}{\sqrt{\omega_m^2 + |\Delta(k \vec{\hat{r}})|^2}} + i\pi \tau_z\right)\right]\\
			&+ E e^{-i\gamma_r} \left[ e^{-\frac{r}{v_F}\sqrt{\omega_m^2 + |\Delta(\theta_{\vec r} +\pi)|^2}} \left(\left(i\omega_m + \Delta(-k \vec{\hat{r}})\tau_+ (i\sigma_y) + {\Delta^*}(-k \vec{\hat{r}})\tau_- (-i\sigma_y)\right)\frac{\pi}{\sqrt{\omega_m^2 + |\Delta(-k \vec{\hat{r}})|^2}} - i\pi \tau_z\right)\right]
		\end{aligned}
	\end{equation}
	where $E = -\sqrt{\frac{k_F}{(2\pi)^3 v_F^2 r}}$. The coefficients of $e^{i\gamma_{\vec r}}$ and $e^{-i\gamma_{\vec r}}$ are $g_{+}(\vec{{r}})$ and $g_{-}(\vec{{r}})$, as defined in the main text (see Eq.10). 
	The phase in the exponential factors $e^{\pm i\gamma_r}$ is defined as $\gamma_r=k_Fr-\pi/4$. This quantity is discussed in the main text beneath Eq.16. The two terms in Eq.\eqref{eq:G_general} describe BdG quasiparticles and quasiholes, propagating along the radius vector $\vec r$. The quantities $\Delta(\theta_{\vec r})$ and $\Delta(\theta_{\vec r} + \pi)$ denote $\Delta(k \vec{\hat{r}})$ and $\Delta(-k \vec{\hat{r}})$, respectively, where $\vec k$ is a vector of length $k_F$ aligned with $\vec r$. 

	
	For a chiral superconductor with winding number $n$, substituting $\Delta(\vec k) = \Delta \left(\frac{k_x + ik_y}{k_F}\right)^n$ into Eq.\eqref{eq:G_general} and simplifying gives
	\begin{equation}
		\begin{aligned}
			G^{(0)}_{d+id}(i\omega_m,\vec r) &= A \tau_0 + B \tau_z + i^n (C \tau_+ + C^\dagger \tau_-),\\
			\begin{pmatrix}
				A \\
				B \\
				C \\
			\end{pmatrix}
			&= D \begin{pmatrix}
				{i\omega_m \cos\phi_r} \\
				-\sin\phi_r \sqrt{\omega_m^2 + \Delta^2} \\
				\Delta (\frac{x+iy}{r})^n \cos(\phi_r - \frac{n\pi}{2})(i\sigma_y)^{q}\\
			\end{pmatrix},
		\end{aligned}
	\end{equation}
	where $D = -\sqrt{\frac{k_F}{2\pi r v_F^2 (\omega_m^2 + \Delta^2)}} e^{-\frac{r}{v_F}\sqrt{\omega_m^2+\Delta^2}}$, 
	$\gamma_r = k_F r - \frac{\pi}{4}$, and $\tau_{\pm} = \frac{\tau_x \pm i \tau_y}{2}$. Here, $q =1$ when $n$ is even and the superconductor is a singlet, and $q=0$ for a spin-polarized triplet superconductor.
	
	For a $d_{x^2-y^2}$ superconductor, substituting $\Delta(\vec k) = \Delta \frac{k_x^2 - k_y^2}{k_F^2}$, and 
	yields 
	\begin{equation}\label{eq:cuprate-zeroth-order-Green's}
		\begin{aligned}
			G^{(0)}_{d_{x^2-y^2}}&(i\omega_m,\vec r) = A_1 \tau_0 + B_1 \tau_z - C_1 \tau_{y} \sigma_y,\\
			\begin{pmatrix}
				A_1 \\
				B_1 \\
				C_1 \\
			\end{pmatrix}
			&= D_1 \begin{pmatrix}
				{i\omega_m \cos\gamma_r} \\
				-\sin\gamma_r \sqrt{\omega_m^2 + \Delta^2 \left(\frac{x^2 - y^2}{r^2}\right)^2} \\
				\Delta \left(\frac{x^2 - y^2}{r^2}\right) \cos\gamma_r\\
			\end{pmatrix},
		\end{aligned}
	\end{equation}
	where $D_1 = -\sqrt{\frac{k_F}{2\pi v_F^2 r\left(\omega_m^2 + \Delta^2 \left(\frac{x^2 - y^2}{r^2}\right)^2\right)}} e^{-\frac{r}{v_F}\sqrt{\omega_m^2 + \Delta^2 \left(\frac{x^2 - y^2}{r^2}\right)^2}}$.

These Green’s functions are used to perturbatively compute the impurity-assisted contribution to 
TDOS. At second order in the perturbation, i.e., in the impurity potential, these contributions yield Eq. 8 of the main text and give rise to the Young-type hyperbolic QPI patterns. The next section explains how Green's functions are used to calculate the TDOS patterns measured in an STM experiment.

\section{Expressing TDOS as a discrete Matsubara sum} 

This section relates the Green's function to the experimentally measured local tunneling density of states (TDOS). 
A standard spectral-function analysis first gives the TDOS as a frequency integral; the finite-temperature TDOS is then rewritten as a sum over discrete Matsubara frequencies.

The thermally averaged TDOS is obtained from the spectral function 
\be\label{eq:local-density-of-states}
A(\omega, \vec r) = -\frac{1}{\pi}\tr\Im\left[G^R_{11}(\omega + i\eta, \vec r, \vec r)\right],
\ee
Here the trace runs over the spin degrees of freedom, and $G_{11}$ denotes the electronic block, which can also be written as $G^R_{11} = \tr \left[(\frac{1+\tau_z}{2}) G^R \right]$, where the trace acts over the particle-hole degrees of freedom.
This quantity is utilized to compute the tunneling current,
\be
I(V, \vec r) = 2 e |t|^2 \nu_0 \int_{-\infty}^\infty d\omega A(\omega, \vec r) [f_{\omega-eV} - f_{\omega }].
\ee

Here $t$ is the STM tip-to-superconductor tunneling amplitude, $\nu_0$ is the density of states at the Fermi level, $f_\omega$ is the Fermi-Dirac distribution function. 
The TDOS $dI/dV$ at zero bias is then given by
\begin{equation}
\left. \frac{dI}{dV} \right|_{V=0}(\vec r)=  \frac{e^2 |t|^2 \nu_0}{2} \int_{-\infty}^\infty d\omega \frac{ A(\omega, \vec r)}{T \cosh^2{(\frac{\omega}{2T})}}
.
\end{equation}
This quantity is nonzero at 
$T>0$ even for a gapped superconductor.
Using the identity 
\[ \frac1{\cosh^{2}(\frac{\omega}{2T})}=\sum_{m}\frac{4 T^2}{(i\omega-\omega_m)^2}
\] 
to perform contour integration over $\omega$, one can rewrite
TDOS ($dI/dV$) at zero bias in a convenient form as
\begin{equation}\label{eq:working-formula-dIdV}
\left. \frac{dI}{dV} \right|_{V=0} 
(\vec r) = 4 e^2 |t|^2 \nu_0 
T\sum_{\omega_m > 0} \tr \Re\left[\left(\frac{1 + \tau_z}{2}\right)\frac{ dG(z=i\omega_m,\vec r,\vec r)}{dz}\right].
\end{equation}

The formula in Eq.\eqref{eq:working-formula-dIdV} can be generalized to finite voltage bias $V$ by considering $\frac{dG}{dz}$ with $z = i\omega_m + eV$, and then summing over the positive Matsubara frequencies.

At zero temperature, $dI/dV$ at finite bias reduces to the simple expression
\begin{equation}\label{eq:dI-dV-T=0}
    \frac{dI}{dV}(\vec r) = 2 e |t|^2 \nu_0 A(eV,\vec r).
\end{equation}

The quantities $\left(\tr\Re [\left(\frac{1 + \tau_z}{2}\right)\frac{ dG(z=i\omega_m,\vec r,\vec r)}{dz}]\right)$ and $\left(\tr\Re[\frac{1}{2}\frac{dG(z=i\omega_m,\vec r,\vec r)}{dz}]\right)$ are identical for any angle-dependent complex gap function $\Delta(\vec k)$ (in other words, $\left(\tr\Im\left[(\frac{1+\tau_z}{2})G^R(\omega + i\eta, \vec r, \vec r)\right]\right)$ and $\left(\tr\Im\left[\frac{1}{2}G^R(\omega + i\eta, \vec r, \vec r)\right]\right)$ are identical). The simpler expression without the additional $(1 + \tau_z)$ is used in the main text. This equivalence follows from particle-hole symmetry: averaging over electron and hole contributions, or retaining only the electronic contribution, produces the same final result.

In this manner, the integral expression of $dI/dV$ is re-expressed as an infinite series over discrete Matsubara frequencies. The resulting series converges rapidly, providing a convenient way to compute the TDOS.

\section{Green's function in a tight binding picture with Wannier functions}

Having considered toy models of the superconductor, the analysis now turns to realistic band structures and crystal lattices. The lattice Green's function of the system can be obtained from the discrete translation-invariant tight-binding Hamiltonian.

For a system with a crystalline lattice, the continuum Green's function can be expressed through a lattice Green's function by dressing it with Wannier functions:
\begin{equation}
G^{(0)}(i\omega_n,\boldsymbol{r},\boldsymbol{r}')=\sum_{\mu,\nu}\sum_{\boldsymbol{R},\boldsymbol{R}'}G^{(0)}_{\mu \nu}(i\omega_n,\boldsymbol{R},\boldsymbol{R}')w_{\mu,\boldsymbol{R}}(\boldsymbol{r})w^*_{\nu,\boldsymbol{R}'}(\boldsymbol{r}') 
\end{equation}
where $\mu,\nu$ are indices in combined Nambu and orbital basis and $\boldsymbol{R},\boldsymbol{R'}$ are lattice vectors in real space.

The second-order correction to Green's function in the presence of two impurities at positions $\boldsymbol{R}_1$ and $\boldsymbol{R}_2$ is given by a sum of two contributions:
\begin{eqnarray}
\begin{aligned}
    G^{(2)}&(i\omega_n,\boldsymbol{r},\boldsymbol{r})
    = \int\int G^{(0)}(i\omega_n,\boldsymbol{r},\boldsymbol{r}_1)V_{\boldsymbol{R}_1}(\boldsymbol{r}_1) G^{(0)}(i\omega_n,\boldsymbol{r}_1,\boldsymbol{r}_2) V_{\boldsymbol{R}_2}(\boldsymbol{r}_2) G^{(0)}(i\omega_n,\boldsymbol{r}_2,\boldsymbol{r}) d^2r_1 d^2 r_2 + \\ &
\int\int G^{(0)}(i\omega_n,\boldsymbol{r},\boldsymbol{r}_1)V_{\boldsymbol{R}_2}(\boldsymbol{r}_1) G^{(0)}(i\omega_n,\boldsymbol{r}_1,\boldsymbol{r}_2) V_{\boldsymbol{R}_1}(\boldsymbol{r}_2) G^{(0)}(i\omega_n,\boldsymbol{r}_2,\boldsymbol{r}) d^2r_1 d^2 r_2
\end{aligned}
\end{eqnarray}
The first term in the equation above can be expressed through the lattice Green's function as follows:
\begin{equation}
\begin{aligned}
\int\int G^{(0)}(i\omega_n,\boldsymbol{r},\boldsymbol{r}_1)V_{\boldsymbol{R}_1}(\boldsymbol{r}_1) G^{(0)}(i\omega_n,\boldsymbol{r}_1,\boldsymbol{r}_2) V_{\boldsymbol{R}_2}(\boldsymbol{r}_2) G^{(0)}(i\omega_n,\boldsymbol{r}_2,\boldsymbol{r}) d^2r_1 d^2 r_2
=
\sum_{\mu\nu \mu_1 \nu_1 \mu_2 \nu_2}
\sum_{\boldsymbol{R}_i \boldsymbol{R}_f \boldsymbol{R}_{a_1}\boldsymbol{R}_{a_2}\boldsymbol{R}_{b_1}\boldsymbol{R}_{b_2}}
 \\
 w_{\mu \boldsymbol{R}_i}(\boldsymbol{r})
G^{(0)}_{\mu \mu_1}(i\omega_n,\boldsymbol{R}_i ,\boldsymbol{R}_{a_1})
\tilde V^{\boldsymbol{R}_1}_{\mu_1 \nu_1}(\boldsymbol{R}_{a_1},\boldsymbol{R}_{b_1})
G^{(0)}_{\nu_1 \nu_2}(i\omega_n,\boldsymbol{R}_{b_1} ,\boldsymbol{R}_{b_2})
\tilde V^{\boldsymbol{R}_2}_{\nu_2\mu_2}(\boldsymbol{R}_{b_2},\boldsymbol{R}_{a_2})
G^{(0)}_{\mu_2 \nu}(i\omega_n,\boldsymbol{R}_{a_2},\boldsymbol{R}_f)
w^*_{\nu \boldsymbol{R}_f}(\boldsymbol{r})
\end{aligned}
\end{equation}
where $\tilde V^{\boldsymbol{R}_0}_{\mu,\nu}(\boldsymbol{R},\boldsymbol{R}')$ is the potential of an impurity at position $\boldsymbol{R}_0$, dressed with Wannier functions:
\begin{eqnarray}
    \tilde V^{\boldsymbol{R}_0}_{\mu,\nu}(\boldsymbol{R},\boldsymbol{R}')=\int w^*_{\mu \boldsymbol{R}}(\boldsymbol{r})V_{\boldsymbol{R}_0}(\boldsymbol{r})w_{\nu \boldsymbol{R}'}(\boldsymbol{r}) d^2r
\end{eqnarray}
For simplicity, point-like impurities are considered, and the Wannier functions are assumed to be well localized with little overlap between different orbitals. In other words, the effective impurity potential is taken to have the form
\begin{eqnarray}
    \tilde V^{\boldsymbol{R}_0}_{\mu,\mu'}(\boldsymbol{R},\boldsymbol{R}')=U_0 \tau^z_{n,n'}\delta_{\sigma \sigma'}\delta_{\boldsymbol{R}_0 \boldsymbol{R}}\delta_{\boldsymbol{R}_0 \boldsymbol{R}'}
\end{eqnarray}
where $n$ and $\sigma$ are the Nambu and orbital components of the index $\mu$. In the absence of interband superconductivity, the expression for the second-order Green's function correction simplifies to
\begin{eqnarray}
\begin{aligned}
    G^{(2)}(i\omega_n,\boldsymbol{r},\boldsymbol{r})
=U_0^2\sum_{\mu \boldsymbol{R}} |w_{\mu \boldsymbol{R}}(\boldsymbol{r})|^2 \tr \big[ G^{(0)}_{\mu\mu}(i\omega_n,\boldsymbol{R},\boldsymbol{R}_1)\tau_z G^{(0)}_{\mu\mu}(i\omega_n,\boldsymbol{R}_1,\boldsymbol{R}_2) \tau_z G^{(0)}_{\mu\mu}(i\omega_n,\boldsymbol{R}_2,\boldsymbol{R})+\\
G^{(0)}_{\mu\mu}(i\omega_n,\boldsymbol{R},\boldsymbol{R}_2)\tau_z G^{(0)}_{\mu\mu}(i\omega_n,\boldsymbol{R}_2,\boldsymbol{R}_1) \tau_z G^{(0)}_{\mu\mu}(i\omega_n,\boldsymbol{R}_1,\boldsymbol{R})
\big]
\end{aligned}
\end{eqnarray}

For the $\gamma$ band, composed of $d_{xy}$ orbitals, a Wannier function with the corresponding orbital structure is used:
\begin{equation}
    w_{xy,\vec R}(\vec r) = \frac{1}{\sqrt{N_{xy}}} (x-R_x) (y-R_y) e^{-\frac{(x-R_x)^2 + (y-R_y)^2}{2\sigma^2}}
\end{equation}
where $N_{xy}$ is the normalization factor, with $\sigma = 0.8 a$.

Similarly, on a particular $x$-$y$ plane, the $d_{xz}$ and $d_{yz}$ orbitals take the form,
\begin{equation}
\begin{aligned}
    w_{xz,\vec R}(\vec r) &= \frac{1}{\sqrt{N_{yz}}} (x-R_x) e^{-\frac{(x-R_x)^2 + (y-R_y)^2}{2\sigma^2}}\\
    w_{yz,\vec R}(\vec r) &= \frac{1}{\sqrt{N_{yz}}} (y-R_y) e^{-\frac{(x-R_x)^2 + (y-R_y)^2}{2\sigma^2}}
\end{aligned}
\end{equation}

The $\mu, \nu$ bands used in the calculations are linear combinations of the $d_{yz}$ and $d_{xz}$ orbitals. In practice, the Wannier functions mainly smear the pattern generated by the lattice Green's function and introduce only small changes; moreover, because the Wannier functions are localized, their convolution is washed out by Fourier filtering. Thus, for the $\alpha$ and $\beta$ bands, an ``averaged" squared Wannier function is used:
\begin{equation}
    |w_{\mu \boldsymbol{R}}(\boldsymbol{r})|^2 = \frac{1}{N} \left[(x-R_x)^2 + (y-R_y)^2 \right] e^{-\frac{(x-R_x)^2 + (y-R_y)^2}{\sigma^2}}
\end{equation}
where $\mu = \alpha, \beta$.

In the main text, the realistic lattice Green's function of strontium ruthenate is modulated by realistic Wannier functions to obtain the real-space propagator. The unperturbed real-space Green's function is then used to determine the second-order contribution to the impurity-modified Green's function, which captures the particle-hole interference effect discussed in the main text.

\section{Details of strontium ruthenate bands and Wannier orbitals}
Strontium Ruthenate is a layered material with three active bands contributing to superconductivity, a nearly circular $\gamma$ band made of $d_{xy}$ orbitals, and two square-like $\alpha$ and $\beta$ bands near the edge of the Brillouin Zone, made of linear combinations of $d_{xz}$ and $d_{yz}$ orbitals. For simplicity, here the analysis considers a model without spin-orbit coupling, as introduced in Refs. \cite{Raghu2012, Wang2013}. The behavior of the normal metal state is captured by the three-band matrix Hamiltonian,
\begin{equation}\label{eq:sr2ruo4-normal-metal}
    H^{(0)} = \sum_{\vec k, \sigma} \Psi^\dagger_{\vec k,\sigma}\begin{pmatrix}
    \epsilon_{xz}(\vec k) & g(\vec k) & 0 \\
    g(\vec k) & \epsilon_{yz}(\vec k) & 0 \\
    0 & 0 & \epsilon_{xy}(\vec k)
\end{pmatrix}\Psi_{\vec k,\sigma}
\end{equation}
where $\Psi_{\vec k, \sigma} = \begin{pmatrix}
    {c_{xz}}_{\vec k, \sigma} & {c_{yz}}_{\vec k, \sigma} & {c_{xy}}_{\vec k, \sigma}
\end{pmatrix}^T$, and
\begin{equation}
    \begin{aligned}
        \epsilon_{xz}(\vec k) &= -2 t_1 \cos(k_x a) - 2 t_2 \cos(k_y a) - \mu\\
        \epsilon_{yz}(\vec k) &= -2 t_2 \cos(k_x a) - 2 t_1 \cos(k_y a) - \mu\\
        g(\vec k) &= -4 t_6 \sin(k_x a) \sin(k_y a) \\
        \epsilon_{xz}(\vec k) &= -2 t_3 (\cos(k_x a) + \cos(k_y a)) 
        - 4 t_4 \cos(k_x a) \cos(k_y a) +\delta - \mu.
    \end{aligned}
\end{equation}
Here, $t_2 = 0.1 t_1$, $t_3 = 0.8 t_1$, $t_4 = 0.35 t_1$, $t_6 = 0.1 t_1$, $\mu = 1.1 t_1$, and $\delta = -0.2 t_1$ (Ref. \cite{Wang2013} and \cite{Raghu2012}) and $t_1 = 0.145$eV \cite{Zabolotnyy2013}. $a$ is the lattice constant. The resulting Fermi surfaces for different bands are shown in Fig. 4(a) in the main text. For the superconducting order parameter, the analysis considers the following order parameter in the band basis $\tilde{\Psi}_{\vec k,\sigma} =\begin{pmatrix}
    {c_{\alpha}}_{\vec k, \sigma} & {c_{\beta}}_{\vec k, \sigma} & {c_{\gamma}}_{\vec k, \sigma}
\end{pmatrix}^T$,
\begin{equation}
    \hat{\Delta}(\vec k) = \begin{pmatrix}
    \Delta(\vec k) & 0 & 0 \\
    0 & \Delta\left(\vec k - \frac{\pi}{a}(\hat{x} + \hat{y})\right) & 0 \\
    0 & 0 & \Delta(\vec k)
\end{pmatrix} \otimes (i\sigma_y).
\end{equation}
For a spin-triplet superconductor, the factor $(i\sigma_y)$ is omitted.

Quasiparticle interference near two impurities from the square-shaped bands, the contribution expected to dominate experimentally, is qualitatively different from its counterpart in the nearly circular $\gamma$ band. In these square-shaped bands, the quasiparticle velocity is concentrated along the $x$ and $y$ axes, suppressing propagation in other directions. By contrast, the two-impurity {QPI} for the realistic $\gamma$-band model is qualitatively similar to the ideal parabolic-band analysis. Thus, {QPI} is shown for both types of bands and for several order-parameter candidates:
\begin{equation}\label{eq:different-gap-functions-sr2ruo4}
    \begin{aligned}
        \Delta_{d+id}(\vec k) &= 2 \Delta_0 (\cos k_y a - \cos k_x a + i \sin k_x a \sin k_y a) \\
        \Delta_{d_{x^2-y^2}}(\vec k) &= 2 \Delta_0 (\cos k_y a - \cos k_x a) \\
        \Delta_{p+ip}(\vec k) &= \Delta_0 (\sin k_x a +i \sin k_y a) \\
        \Delta_{s}(\vec k) &= \Delta_0 .
    \end{aligned}
\end{equation}

To accurately capture the spatial variation of tunneling conductance on the sublattice scale, the BdG+W method is employed \cite{ChoubeyBdGW2014, kreiselBdGW2015} where the continuum Green's function can be expressed through a lattice Green's function by convoluting it with the Wannier functions:
\begin{equation}\label{eq:wannier-modulated-greens-fn}
G^{(0)}(i\omega_n,\vec{r},\vec{r}')=\sum_{\mu,\nu}\sum_{\vec{R},\vec{R}'}G^{(0)}_{\mu \nu}(i\omega_n,\vec{R},\vec{R}')w_{\mu,\vec{R}}(\vec{r})w^*_{\nu,\vec{R}'}(\vec{r}'),
\end{equation}
where $w_{\mu \vec R}(\vec r)$ denotes the Wannier function of orbital $\mu$ centered at lattice point $\vec R$, evaluated at the continuum point $\vec r$. The lattice Green's function is computed numerically from the BdG Hamiltonian and substituted into Eq. 8 of the main text and Eq. \eqref{eq:working-formula-dIdV} to obtain the local TDOS patterns.

The difference between this approach and a conventional BdG analysis on a lattice is illustrated in Fig. 6 in the main text, where the joint contribution to TDOS due to two impurities has been plotted, while taking into account the effect of the Wannier functions. In both cases, only the two-impurity interference contribution is plotted after removing Friedel oscillations.

It has been argued that the out-of-plane nature of the orbitals in the $\alpha, \beta$ bands greatly enhances their tunneling amplitude to the STM tip, compared to the in-plane $\gamma$ band \cite{kreisel2021}. Therefore, the quasiparticle interference in the $\alpha, \beta$ bands will dominate the experimentally measured tunneling conductance signal.

\section{Tunneling conductance of circular bands in strontium ruthenate}

\begin{figure}[h]
    \centering
    \includegraphics[width=0.99\linewidth]{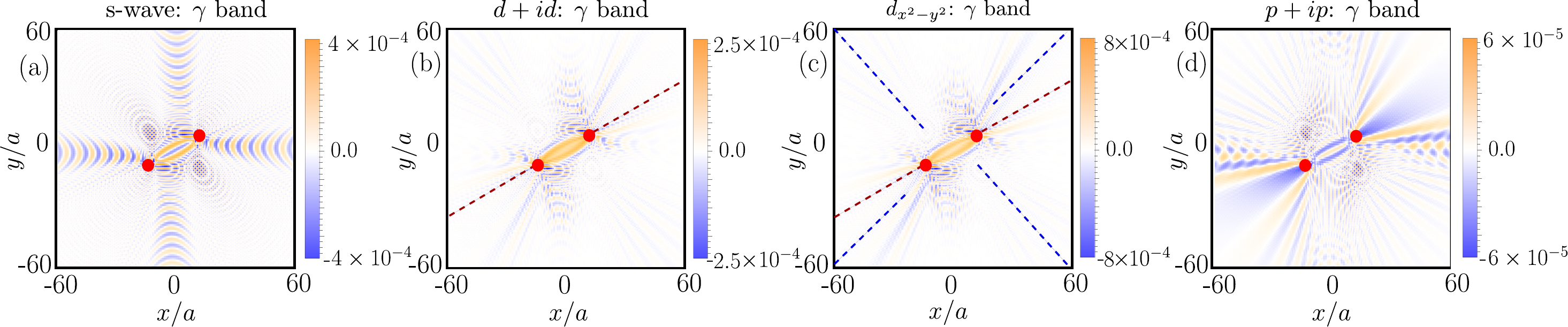}
    \caption{Second-order contribution to $\frac{dI}{dV}$ for various order parameters in Sr$_2$RuO$_4$: The panels show the contribution to $dI/dV$ from the $\gamma$ band (including the effect of Wannier orbitals) for (a) $s$-wave , (b) $d+id$ (c) $d_{x^2-y^2}$ and (d) $p+ip$ superconductor. TDOS is plotted in units of  $\frac{4 e^2 |t_{\rm STM}|^2 \nu_0 U_0^2}{(t_1 a^2)^3}$, for a configuration where the impurities at distance $30 a$ are rotated by angle $30^\circ$ with respect to the $x$ axis. Note that the tunneling amplitude $t_{\rm STM}$ to STM is much smaller for the $\gamma$ band compared to that of $\alpha, \beta$ bands depicted in the main text. 
    The spin-triplet $p+ip$ superconductor (panel (d)) has a distinct feature that there is a strong second-order interference effect on the line joining the impurities when the impurities are not aligned with the axes of the lattice. This feature is also observed in an ideal parabolic band structure.
    The gap amplitude is set to $\Delta_0 = t_1/100$ (see Eq. \eqref{eq:different-gap-functions-sr2ruo4}), and the temperature is $k_B T = 0.5 \Delta_0$.}
    \label{fig:gamma_unfiltered_sr2ruo4}
\end{figure}

Figure~\ref{fig:gamma_unfiltered_sr2ruo4} shows the TDOS near two impurities for the $\gamma$ band of strontium ruthenate. While the circular $\gamma$ band is not the most important band, this band provides a useful consistency check.
The behavior of TDOS in these bands is qualitatively similar to that of an ideal parabolic band in the continuum limit.

Strontium ruthenate also has square-shaped $\alpha$ and $\beta$ bands. Due to their out-of-plane orbital character, they strongly couple to the STM tip, and their contributions to TDOS are discussed in the main text.

\newpage
\section{Fourier filtering of TDOS: Isolation of Young's interference fringes and  nodal lines}

The experimental TDOS pattern near two impurities will be dominated by the Friedel oscillation of individual impurities.
This section discusses how Fourier filtering can enhance the visibility of hyperbolic particle-hole interference patterns and reveal their nodal lines.
The separation of particle-hole and particle-particle contributions proves feasible because Friedel oscillations have wavenumbers $\approx 2k_F$, whereas the particle-hole 
contribution, being essentially Young's-interference, occurs at markedly lower wavenumbers $k\lesssim 1/|\vec r_1-\vec r_2|$. These characteristic wavelengths are functions of the band structure and the geometric configuration of the impurities, and their presence, as well as their nodal lines, do not depend on the energy bias at which they are probed, as long as the energy scale is not too large compared to $\Delta$. 

\begin{figure}[h]
    \centering
    \includegraphics[width=0.5\linewidth]{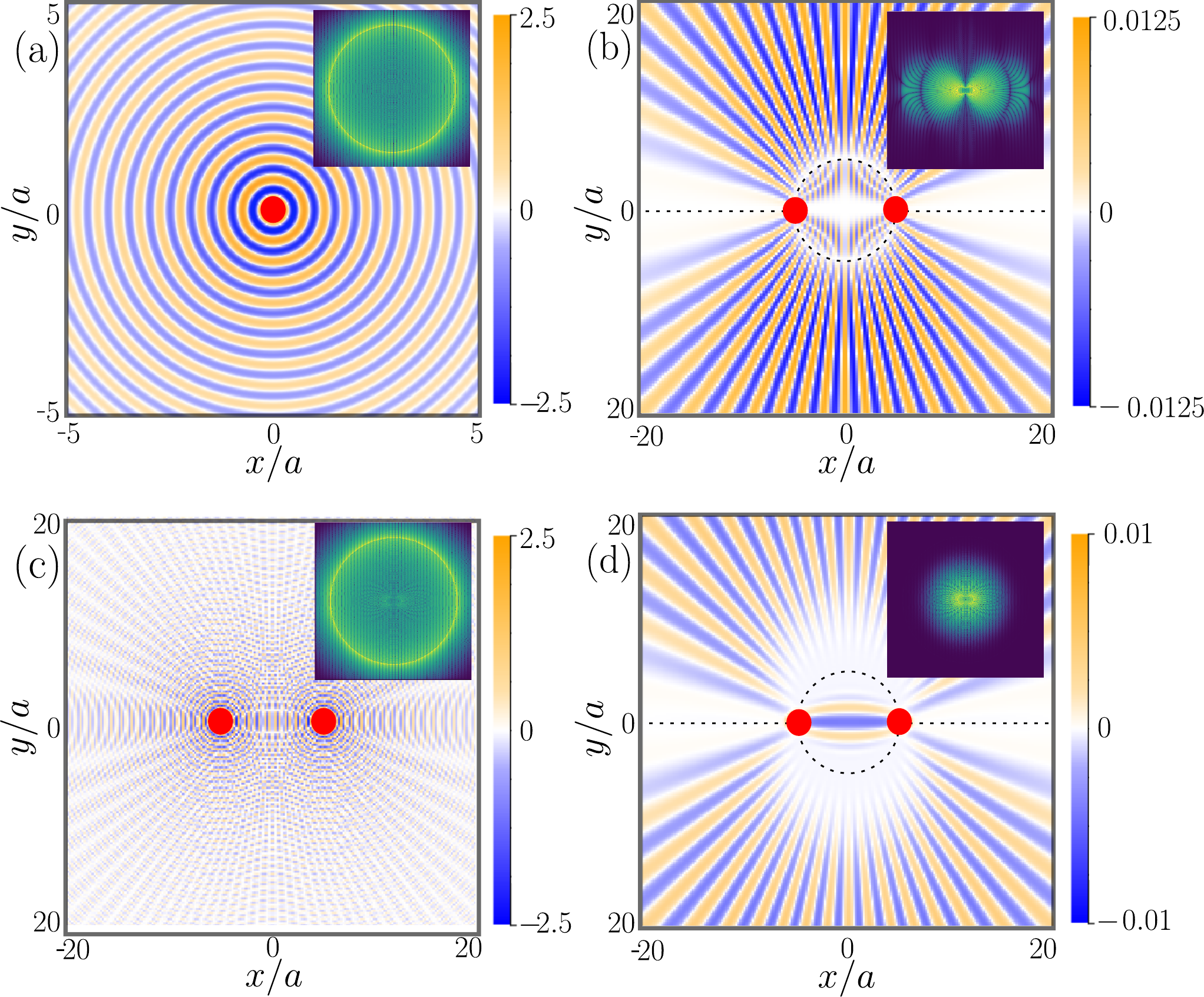}
    \caption{(a) Friedel oscillation in TDOS near a single impurity on a $d+id$ superconductor. The concentric rings spaced by distance $\lambda_F/2$ correspond to a peak at wavenumber $2 k_F$, which is evident from the Fourier transform of the pattern, shown in the inset (with a logarithmic color scheme). For simplicity, the spatial patterns of $dI/dV$ are shown at zero temperature and a bias voltage of $eV = 1.1 \Delta$ (see Eq.\eqref{eq:dI-dV-T=0}). The TDOS is plotted in the units of $4 e^2 |t_{\text{STM}}|^2 \nu_0 U_0  \left(\frac{k_F}{2\pi \lambda_F v_F^2}\right)$. (b) The spatial pattern of the hyperbolic Young's interference fringes, with its Fourier transform shown in the inset. Here, the peak in the Fourier space occurs at a wavenumber much smaller than $2 k_F$. The dimensionless impurity strength is $U_0 \sqrt{\frac{k_F}{2\pi \lambda_F v_F^2}} = 0.01$. (c) The total TDOS for two impurities is shown. Here, the spatial pattern is dominated by Friedel oscillations from the two impurities, which is evident from the Fourier transform of the pattern shown in the inset. (d) The Fourier filtered pattern which displays hyperbolic fringes. For that the large-$k$ harmonics from the Fourier-space pattern in panel (c) inset were removed with a smooth Gaussian filter, and the resulting Fourier pattern was inverse transformed to real space.}
    \label{fig:FFT-filtering}
\end{figure}

Fourier filtering of Young's interference fringes is illustrated in Fig. \ref{fig:FFT-filtering}. The original Fourier-space data are multiplied by $e^{-{(k_x^2 + k_y^2)}/{(2 \sigma^2)}}$, with $\sigma=2k_F$, to suppress contributions from large-$k$ harmonics, and are then inverse-transformed back to real space. This filtering makes the radial Young's interference patterns more prominent while attenuating the Friedel oscillations.

\begin{figure}[h]
\centering
\includegraphics[width=0.57\linewidth]{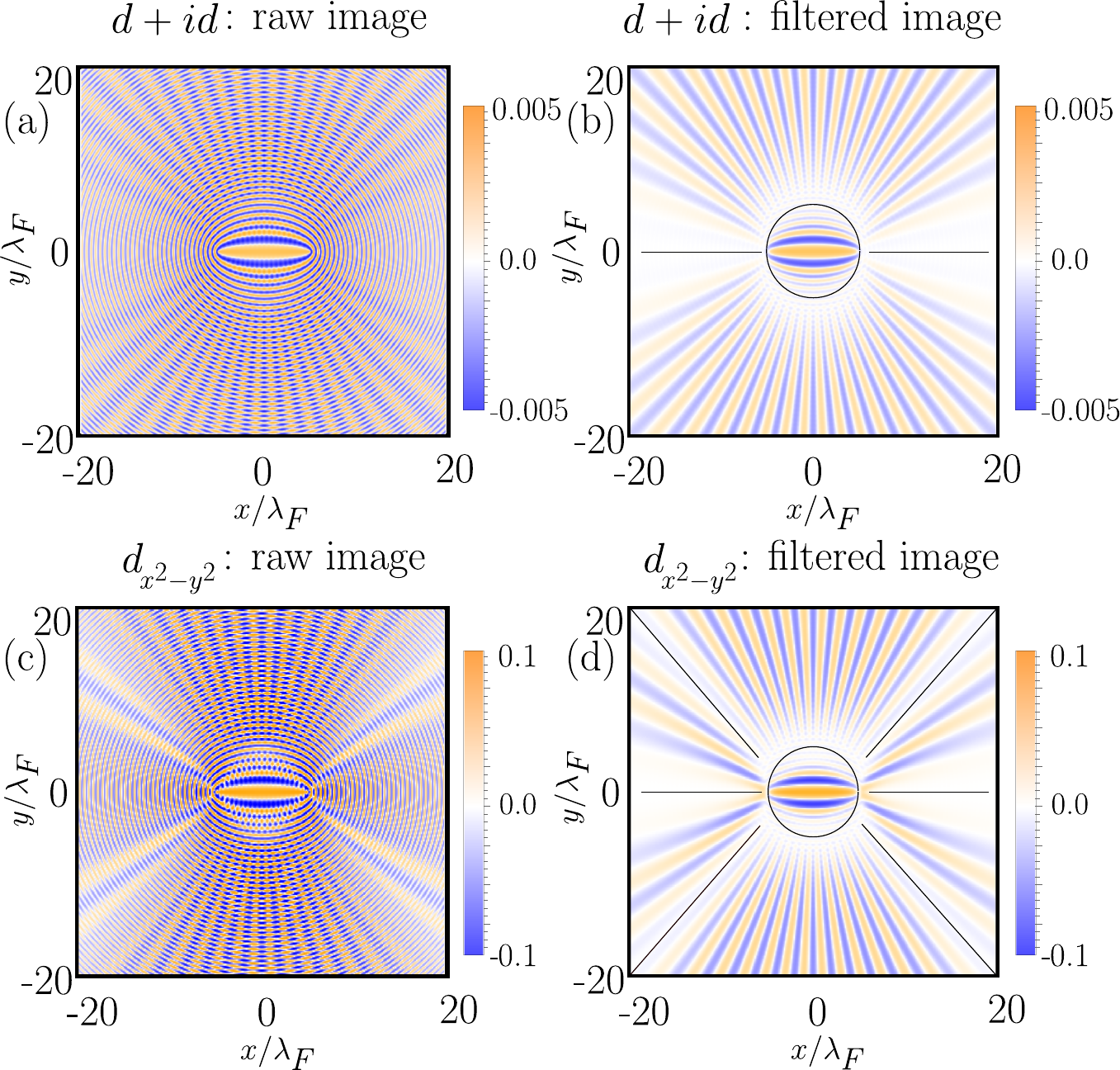}
\caption{ Enhancing the visibility of nodal lines in particle-hole interference by Fourier filtering illustrated for a $d+id$ superconductor (a, b) and a nodal $d$-wave ($d_{x^2 - y^2}$) superconductor (c, d). The two-impurity contribution to the TDOS 
includes both particle-particle and particle-hole interference terms, with the former obscuring the nodal features of the latter at distances $r\sim |\vec r_1-\vec r_2|$.
After applying Fourier filtering (see text), the nodal lines in the particle-hole interference become clearly visible, matching quite well the nodal lines in the particle-hole interference seen in Fig. 1 of the main text (thin black lines). Here the parameters are $k_B T = 0.5 \Delta$ in (a,b) and $k_B T = 0.2 \Delta$ in (c,d). Here, TDOS is measured in the units of $4 e^2 |t_{\text{STM}}|^2 \nu_0 U_0^2  \left(\frac{k_F}{2\pi \lambda_F v_F^2}\right)^{3/2}$.
}
\label{fig:IFFT}
\end{figure}

Filtering also reveals the nodal beams in the Young's interference fringes---lines along which the fringe contrast is depleted. As discussed in the main text, these nodal beams contain information about the angular dependence and topology of the gap function. This is illustrated in Figs. \ref{fig:FFT-filtering} and \ref{fig:IFFT}, which show Fourier filtering for $d+id$ and $d_{x^2-y^2}$ superconductors. In both cases, the filtering preserves the static and rotating nodal beams that are central to the SOPT method. The resulting nodal lines agree well with those in the particle-hole interference contributions shown in Fig. 1 of the main text.

%
%

